\documentclass[journal]{IEEEtran}

\usepackage{cite}

\ifCLASSINFOpdf
   \usepackage[pdftex]{graphicx}
   \graphicspath{{../pdf/}{../jpeg/}}
   \DeclareGraphicsExtensions{.pdf,.jpeg,.png}
\else
   \usepackage[dvips]{graphicx}
   \graphicspath{{../eps/}}
   \DeclareGraphicsExtensions{.eps}
\fi

\usepackage{amsmath,bm}

\usepackage{algorithmic}

\usepackage{array}

\ifCLASSOPTIONcompsoc
  \usepackage[caption=false,font=normalsize,labelfont=sf,textfont=sf]{subfig}
\else
  \usepackage[caption=false,font=footnotesize]{subfig}
\fi

\usepackage{fixltx2e}

\usepackage{url}

\usepackage{soul}
\usepackage{color}
\usepackage{setspace}
\usepackage{multirow}
\usepackage[flushleft]{threeparttable}

\usepackage{bm}
\usepackage{amsbsy}
\usepackage{float}
\usepackage{mathtools}

\makeatletter
\def\endthebibliography{%
  \def\@noitemerr{\@latex@warning{Empty `thebibliography' environment}}%
  \endlist
}
\makeatother

\begin{document}

\setstcolor{red}

\newcommand{\etal}{\textit{et al}. }
\newcommand{\ie}{\textit{i}.\textit{e}., }
\newcommand{\eg}{\textit{e}.\textit{g}. }

\newcommand{\matJ}[0]{\widetilde{\mathbf{J}}}
\newcommand{\matM}[0]{\mathbf{M}}
\newcommand{\matX}[0]{\mathbf{X}}
\newcommand{\matE}[0]{\mathbf{E}}
\newcommand{\matH}[0]{\widetilde{\mathbf{H}}}

\newcommand{\opLE}[0]{\mathcal{L}_E}
\newcommand{\opLH}[0]{\mathcal{L}_H}
\newcommand{\opK}[0]{\mathcal{K}}
\newcommand{\opKpv}[0]{\widetilde{\mathcal{K}}}

\newcommand{\matf}[0]{\mathbf{f}}
\newcommand{\matPhi}[0]{\mathbf{\Phi}}

\newcommand{\matL}[0]{\mathbf{L}}
\newcommand{\matK}[0]{\mathbf{K}}

\newcommand{\nx}[0]{\hat{n} \times}
\newcommand{\nxnx}[0]{\hat{n} \times \hat{n} \times}

\newcommand{\vect}[1]{\boldsymbol{#1}}
\newcommand{\matr}[1]{\mathbf{#1}}
\newcommand{\vr}[0]{\vect{r}}

\newcommand{\fE}[0]{\matr{E}}
\newcommand{\fH}[0]{\matr{H}}

\newcommand{\IntOmega}[0]{\int_{\omega_0 - \Delta \omega}^{\omega_0 + \Delta \omega}}

\title{Wigner-Smith Time Delay Matrix for Electromagnetics: Guiding and Periodic Systems with Evanescent Modes}

\author{Yiqian~Mao,
	  Utkarsh~R.~Patel
        and~Eric~Michielssen*,~\IEEEmembership{Fellow,~IEEE}
\thanks{Y. Mao, U. R. Patel and E. Michielssen are with the Department of Electrical Engineering and Computer Science, University of Michigan, Ann Arbor, MI 48109 USA. (e-mail: yqmao@umich.edu; utkarsh.patel@alum.utoronto.ca; emichiel@umich.edu).}
}


\maketitle

\begin{abstract}
The Wigner-Smith (WS) time delay matrix relates an electromagnetic system's scattering matrix and its frequency derivative. Previous work showed that the entries of WS time delay matrices of systems excited by propagating waves consist of volume integrals of energy-like field quantities. This paper introduces a generalized WS relationship that applies to systems excited by mixtures of propagating and evanescent fields. Just like its predecessor, the generalized WS relationship allows for the identification of so-called WS modes that interact with the system with well-defined time delays. Furthermore, a technique is developed to compute the WS time delay matrix of a composite system from the WS time delay matrices of its subsystems. Numerical examples demonstrate the usefulness of the generalized WS method when characterizing time delays experienced by fields interacting with guiding and periodic structures that have ports supporting evanescent modes. 
\end{abstract}

\begin{IEEEkeywords}
Wigner-Smith time delay, generalized S-matrix, evanescent mode, cascade system.
\end{IEEEkeywords}

\IEEEpeerreviewmaketitle

\section{Introduction}
\label{sec:intro}

\IEEEPARstart{T}{he} Wigner-Smith (WS) time delay matrix
\begin{align}
\label{eq:Q_def1}
\matr{Q} = j \matr{S}^\dag \frac{\partial \matr{S}}{\partial \omega} \,,
\end{align}
was introduced by Felix Smith to characterize a particle's dwell time in a quantum system with scattering matrix $\matr{S}$ \cite{Smith1960Lifetime}. Today, the WS time delay matrix is used in acoustics, optics, and electromagnetics to characterize group delays of wave packets \cite{Gerard2016part}, to synthesize modes that travel through systems with minimal dispersion \cite{Carpen2015Obs}, to shape waves traveling through disordered media  \cite{Hougne2021coherent, Bohm2016microwave}, and to determine frequency sensitivities of antenna input impedances \cite{Patel2020WS1, Patel2020WS2}. More recent applications can be found in \cite{Gallmann_2017, Brandstotter_2019, Gerardin2014full, Bohm_2018, Durand_2019, Texier_2016, Hougne2021demand, Chen2021general}. 

Patel and Michielssen recently developed techniques to directly compute $\matr{Q}$ for 3-D electromagnetic systems \emph{excited by propagating modes} using volume integrals of energy-like field quantities \cite{Patel2020WS1, Patel2020WS2}. Their methods allow for the finite-element or integral equation-based evaluation of $\matr{Q}$ and its subsequent diagonalization to construct WS modes that exhibit well-defined group delays. Knowledge of $\matr{Q}$ and $\matr{S}$ also allows for the computation of the frequency derivative of $\matr{S}$, which in turn may advance the state of the art in fast frequency-sweep methods.

Unfortunately, the methods in  \cite{Patel2020WS1, Patel2020WS2} do not apply to systems excited by mixtures of propagating and evanescent modes. This limits their applicability to real-world problems involving structures with large multimode apertures that reside near structural discontinuities and/or material inhomogeneities. 

The restrictions of the methods in \cite{Patel2020WS1, Patel2020WS2} at times can be sidestepped by shifting ports so they no longer contain evanescent fields. Unfortunately, this fix artificially enlarges the system, increasing the cost of computing $\matr{Q}$. 

The restrictions of the methods in \cite{Patel2020WS1, Patel2020WS2} also limit their application to the closed-loop design of composite systems with prescribed time-delays. Such systems oftentimes consist of multiple components/subsystems that require iterative refinement. The straightforward computation of the composite system's $\matr{Q}$ requires usually is prohibitively expensive, especially if the refinement only involves a small set of subsystems. A better strategy is to compute $\matr{Q}$ of each component and combine them to obtain the $\matr{Q}$ of the composite system. Using this approach only $\matr{Q}$'s of the components being modified require updating. This approach however requires that the system can be flexibly partitioned, oftentimes introducing ports that support evanescent modes.

This paper extends the methods of \cite{Patel2020WS1} to systems excited by mixtures of propagating and evanescent modes. Its contributions are five-fold manner:
\begin{enumerate}
\item It generalizes WS relationship ~\eqref{eq:Q_def1}, valid for systems excited by propagating waves, to one applicable to systems fed by mixtures of propagating and evanescent waves; both guiding and periodic systems are considered.
\item It demonstrates that the generalized WS relationship allows the frequency sensitivities of scattering parameters to be evaluated without inverting the scattering matrix.   
\item It illustrates that the generalized WS relationship permits the identification of so-called WS modes that experience well-defined time delays when interacting with the system. These WS modes are constructed by diagonalizing the portion of the WS time delay matrix modeling propagating modes.
\item It shows that the generalized WS time delay matrix of a composite system can be computed from the WS time delay matrices of its subsystems.
\item It numerically demonstrates the applicability of the generalized WS scheme to guiding and periodic structures with ports supporting evanescent modes. 
\end{enumerate}

The above contributions remove all the aforementioned restrictions of the methods in \cite{Patel2020WS1, Patel2020WS2}.  Contributions 1-4 and 5 are detailed in Sections~\ref{sec:em} and \ref{sec:example} below. Throughout this paper, $f$ denotes frequency and $\omega = 2 \pi f$ denotes angular frequency.  In addition, $'$ denotes $\partial/\partial \omega$, and $^*$, $^T$, and $^\dag$ denote conjugate, transpose, and conjugate transpose operations.

\section{WS Relationship Including Evanescent Modes}
\label{sec:em}

\subsection{Generalized WS Relationship for Guiding Systems}
\label{sec:em_guiding}

This section generalizes WS relationship ~\eqref{eq:Q_def1} to systems fed by mixtures of propagating and evanescent waves. The elements of the generalized WS time delay matrix are shown to consist of volume integrals of energy-like densities plus correction terms that account for the propagating/evanescent character of the system modes.

\begin{figure}[hbtp!]
\centering\includegraphics[width=6.0cm]{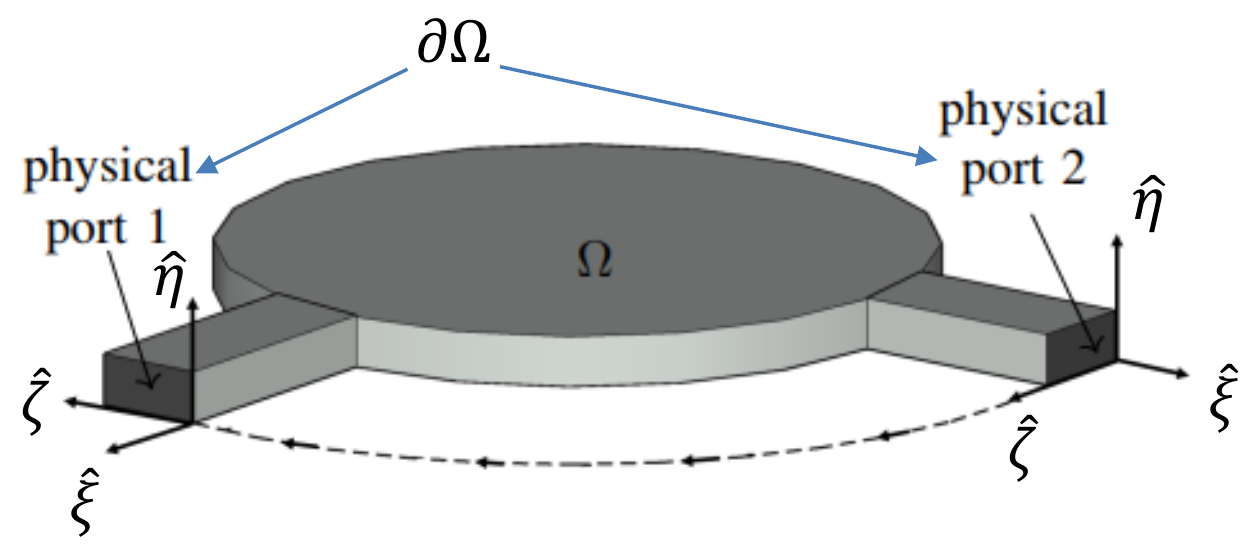}
\caption{Guiding system under consideration \cite{Patel2020WS1}.}
\label{fig:Guiding_illus}
\end{figure}

Consider a guiding system composed of perfect electrically conducting (PEC) cavities and waveguides that occupy volume $\Omega$ (Fig.~\ref{fig:Guiding_illus}). Let $\partial \Omega$ denote the union of all physical waveguide ports. Assume the vicinity of $\partial \Omega$ is parameterized by a locally Cartesian coordinate system $(\xi,\eta,\zeta)$ with $\zeta > 0$ exterior to $\Omega$. Assume that $\Omega$ is filled with a nondispersive and lossless material with permittivity $\varepsilon(\vr)$ and permeability $\mu(\vr)$, which are constant for $\vr$ near $\partial \Omega$. Under these conditions, fields that enter and exit $\Omega$ can be expanded into modes with real and frequency-independent transverse profiles $\bm{\mathcal{X}}_p(\eta,\zeta)$, propagation constants $\beta_p$, and impedances $Z_p$, $p=1, ..., M$ (see Appendix A in \cite{Patel2020WS1}). 
In contrast to the assumptions underlying the study of WS methods in \cite{Patel2020WS1}, some of these modes may be evanescent. For propagating (evanescent) modes, $\beta_p$ and $Z_p$ are purely real (imaginary). The mode profiles satisfy the orthonormality condition
\begin{align}
\int_{\partial \Omega} \bm{\mathcal{X}}_p(\eta,\zeta) \cdot \bm{\mathcal{X}}_q(\eta,\zeta) d\eta d\zeta = \delta_{pq} \,, \label{eq:Guiding_orthogonal}
\end{align}
where $\delta_{pq}$ is the Kronecker delta.

Assume the system is excited by the unit-power incoming wave with transverse field components near $\partial \Omega$ given by
\begin{subequations}
\begin{align}
\fE_{p,\|}^i (\vr,\omega) &= n_p e^{j \beta_{p}(\omega) \xi} \bm{\mathcal{X}}_p(\eta,\zeta) \label{eq:incoming_E_transv} \\
\fH_{p,\|}^i (\vr,\omega) &= \frac{1}{n_p} e^{j \beta_{p}(\omega) \xi} \left(- \hat{\xi} \times \bm{\mathcal{X}}_p(\eta,\zeta)\right) \,, \label{eq:incoming_H_transv}
\end{align}
\end{subequations}
where $n_p = \sqrt{Z_p}$. In what follows, spatial and frequency dependencies are omitted when possible. The outgoing transverse fields near $\partial \Omega$ in response to this excitation are 
\begin{subequations}
\begin{align}
\fE_{p,\|}^o (\vr,\omega) &= \sum_{m=1}^{M} S_{mp} n_m e^{- j \beta_{m} \xi} \bm{\mathcal{X}}_m \\
\fH_{p,\|}^o (\vr,\omega) &= - \sum_{m=1}^{M} S_{mp} \frac{1}{n_m} e^{- j \beta_{m} \xi} (- \hat{\xi} \times \bm{\mathcal{X}}_m)
\end{align}
\end{subequations}
where $S_{mp}$ is the scattering coefficient expressing coupling from mode $p$ to mode $m$. For $\vr$ near $\partial \Omega$, the total transverse fields are
\begin{subequations}
\begin{align}
\fE_{p,\|} &= \fE_{p,\|}^i + \fE_{p,\|}^o \label{eq:tot_E_transv} \\
\fH_{p,\|} &= \fH_{p,\|}^i + \fH_{p,\|}^o \label{eq:tot_H_transv} \,.
\end{align}
\end{subequations}

Let \{$\fE_{p/q}(\vr,\omega), \fH_{p/q}(\vr,\omega)$\} denote total fields excited by mode $p/q$ for all $\vr \in \Omega$. 
Taking the frequency derivative of Maxwell's equations for \{$\fE_p, \fH_p$\} while conjugating those for \{$\fE_q, \fH_q$\} yields
\begin{subequations}
\begin{align}
\nabla \times \fE_p' &= -j \omega \mu \fH_p' - j \mu \fH_p \label{eq:dw_nabla_Ep} \\
\nabla \times \fH_p' &= j \omega \varepsilon \fE_p' + j \varepsilon \fE_p \label{eq:dw_nabla_Hp} \\
\nabla \times \fE_q^* &= j \omega \mu \fH_q^* \label{eq:conj_nabla_Eq} \\
\nabla \times \fH_q^* &= - j \omega \varepsilon \fE_q^* \,. \label{eq:conj_nabla_Hq}
\end{align}
\end{subequations}
The dot product of Eq.~\eqref{eq:dw_nabla_Hp} and $\frac{-j}{2} \fE_q^*$ plus the dot product of Eq.~\eqref{eq:conj_nabla_Hq} and $\frac{-j}{2} \fE_p'$ reads
\begin{align}
- \frac{j}{2} \fE_p' \cdot \nabla \times \fH_q^* - \frac{j}{2} \fE_q^* \cdot \nabla \times \fH_p' = \frac{1}{2} \varepsilon \fE_q^* \cdot \fE_p \,. \label{eq:interm_1}
\end{align}
Similarly, the dot product of Eq.~\eqref{eq:dw_nabla_Ep} and $\frac{-j}{2} \fH_q^*$ plus the dot product of Eq.~\eqref{eq:conj_nabla_Eq} and $\frac{-j}{2} \fH_p'$ reads
\begin{align}
\frac{j}{2} \fH_p' \cdot \nabla \times \fE_q^* + \frac{j}{2} \fH_q^* \cdot \nabla \times \fE_p' = \frac{1}{2} \mu \fH_q^* \cdot \fH_p \,. \label{eq:interm_2}
\end{align}
Subtracting Eq.~\eqref{eq:interm_2} from Eq.~\eqref{eq:interm_1} produces
\begin{align}
\frac{j}{2} \nabla \cdot \left( \fE_p' \times \fH_q^* + \fE_q^* \times \fH_p' \right) = \frac{1}{2} \left( \varepsilon \fE_q^* \cdot \fE_p + \mu \fH_q^* \cdot \fH_p \right) \,. \label{eq:Helmholtz_pre_integ}
\end{align}
Next, integrating the left- and right-hand side (LHS and RHS) of  Eq.~\eqref{eq:Helmholtz_pre_integ} over $\Omega$ and applying the divergence theorem yields
\begin{align}
&\frac{j}{2} \int_{\partial \Omega} \hat{\xi} \cdot \left( \fE_{p,\|}' \times \fH_{q,\|}^* + \fE_{q,\|}^* \times \fH_{p,\|}' \right) dS \nonumber \\
& \quad \quad = \frac{1}{2} \int_{\Omega} \left( \varepsilon \fE_q^* \cdot \fE_p + \mu \fH_q^* \cdot \fH_p \right) dV \,\label{eq:Helmholtz_g_integ2}
\end{align}
where use was made of the fact that surface integrals involving electric fields tangential to the system's PEC walls vanish. 
A lengthy but straightforward evaluation of the LHS of Eq.~\eqref{eq:Helmholtz_g_integ2} using Eqs.~\eqref{eq:incoming_E_transv}--\eqref{eq:tot_H_transv} yields
\begin{align}
& \int_{\partial \Omega} \hat{\xi} \cdot \left( \fE_{p,\|}' \times \fH_{q,\|}^* + \fE_{q,\|}^* \times \fH_{p,\|}' \right) dS \nonumber \\
& \quad = - \delta_{pq} \left[ \frac{n_p'}{n_q^*} + n_q^* \left( \frac{1}{n_p} \right)' \right] + S_{pq}^* \left[ \frac{n_p'}{n_p^*} - n_p^* \left( \frac{1}{n_p} \right)' \right] \nonumber \\
& - S_{qp} \left[ \frac{n_q'}{n_q^*} - n_q^* \left( \frac{1}{n_q} \right)' \right] + \sum_{m=1}^M S_{mq}^* S_{mp}' \left( \frac{n_m}{n_m^*} + \frac{n_m^*}{n_m} \right) \nonumber \\
& - S_{qp}' \left( \frac{n_q}{n_q^*} - \frac{n_q^*}{n_q} \right) + \sum_{m=1}^M S_{mq}^* S_{mp} \left[ \frac{n_m'}{n_m^*} + n_m^* \left( \frac{1}{n_m} \right)' \right] \,. \nonumber
\end{align}
Upon defining
\begin{align}
\nu_{\pm,p} &= \frac{j}{2} \left( \frac{n_p^*}{n_p} \pm \frac{n_p}{n_p^*} \right) \,, \label{eq:def_rho_pm}
\end{align}
\begin{align}
\lambda_{\pm,p} &= \frac{j}{2} \left[ n_p^* \left( \frac{1}{n_p} \right)' \pm n_p' \frac{1}{n_p^*} \right] \,, \label{eq:def_lambda_pm}
\end{align}
and 
\begin{align}
\widetilde{Q}_{qp} = \frac{1}{2} \int_{\Omega} \left( \varepsilon \fE_q^* \cdot \fE_p + \mu \fH_q^* \cdot \fH_p \right) dV \,, \label{eq:Q_tilde}
\end{align}
Eq.~\eqref{eq:Helmholtz_g_integ2} simplifies to
\begin{align}
\widetilde{Q}_{qp} &= - \delta_{pq} \lambda_{+,p} - S_{pq}^* \lambda_{-,p} + S_{qp} \lambda_{-,q} + \sum_m S_{mq}^* S_{mp}' \nu_{+,m} \nonumber \\
& \quad + S_{qp}' \nu_{-,q} + \sum_m S_{mq}^* S_{mp} \lambda_{+,m} \,. \label{eq:LHS_surf_integ_result2}
\end{align}
To compactly express Eq.~\eqref{eq:LHS_surf_integ_result2} in matrix form, let $\matr{N}_{\pm}$ and $\matr{\Lambda}_{\pm}$ denote diagonal matrices with elements
\begin{align}
\left( \matr{N}_{\pm} \right)_{pp} &= \nu_{\pm,p} \label{eq:Nu} \\
\left( \matr{\Lambda}_{\pm} \right)_{pp} &= \lambda_{\pm,p} \,. \label{eq:Lambda}
\end{align}
Substituting Eqs.~\eqref{eq:Nu}--\eqref{eq:Lambda} into Eq.~\eqref{eq:LHS_surf_integ_result2} yields
\begin{align}
\matr{Q} = \left( \matr{N}_{-} +  \matr{S}^\dag \matr{N}_{+} \right) \matr{S}'  \label{eq:WS_gen2}
\end{align}
where
\begin{align}
\matr{Q} = \widetilde{\mathbf{Q}} + \matr{\Lambda}_{+} + \mathbf{S}^\dag \matr{\Lambda}_{-} - \matr{\Lambda}_{-} \matr{S} - \matr{S}^\dag \matr{\Lambda}_{+} \mathbf{S} \,. \label{eq:volumeQ_plus_correction}
\end{align}
Eqs.~\eqref{eq:WS_gen2} and \eqref{eq:volumeQ_plus_correction} constitute the sought generalized WS relationship involving both propagating and evanescent modes. Eq.~\eqref{eq:WS_gen2} is therefore a generalization of Eq.~\eqref{eq:Q_def1}.

In the absence of evanescent modes, \ie when only propagating modes are present in $\partial \Omega$, Eqs.~\eqref{eq:def_rho_pm} and \eqref{eq:def_lambda_pm} simplify to
\begin{align}
\nu_{+,p} &= j \label{eq:nu_plus_p_prop} \\
\nu_{-,p} &= \lambda_{+,p} = 0 \\
\lambda_{-,p} &= j n_p \left( \frac{1}{n_p} \right)' \,. \label{eq:lambda_minus_p_prop}
\end{align}
Substituting Eqs.~\eqref{eq:nu_plus_p_prop}--\eqref{eq:lambda_minus_p_prop} into Eqs.~\eqref{eq:Nu}--\eqref{eq:Lambda} simplifies Eq.~\eqref{eq:WS_gen2} to Eq.~\eqref{eq:Q_def1}, the WS relationship derived in \cite{Smith1960Lifetime}.

\subsection{Generalized WS Relationship for Periodic Systems}
\label{sec:em_Floquet}

This section extends the formulation of Section II.A to doubly periodic systems subject to propagating and evanescent Floquet mode excitations.

\begin{figure*}[htbp!]
\centering\subfloat[ \label{fig:Prd_illus1}]{\centering\includegraphics[width=0.7\columnwidth]{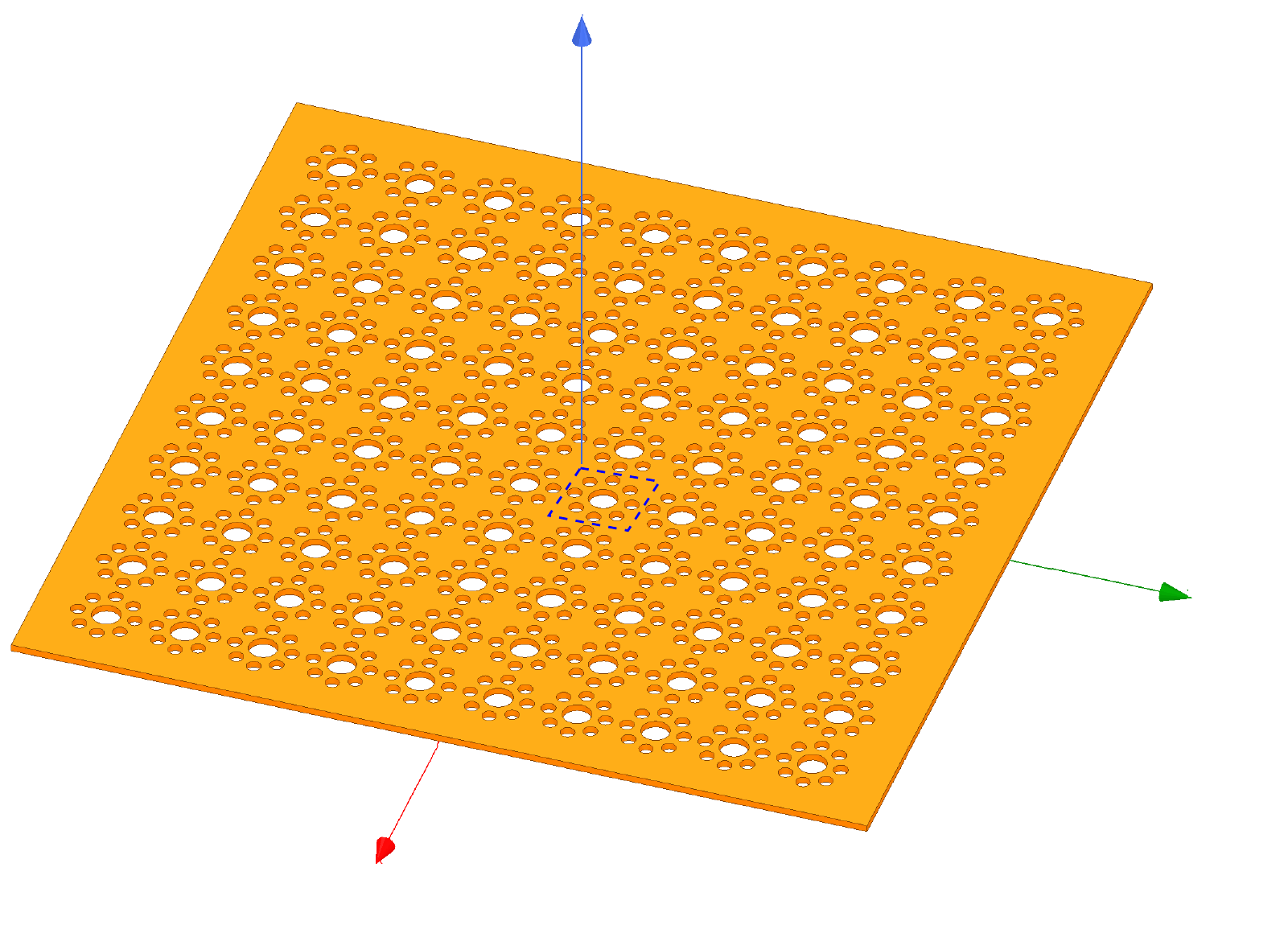}} \qquad
\centering\subfloat[ \label{fig:Prd_illus2}]{\centering\includegraphics[width=0.7\columnwidth]{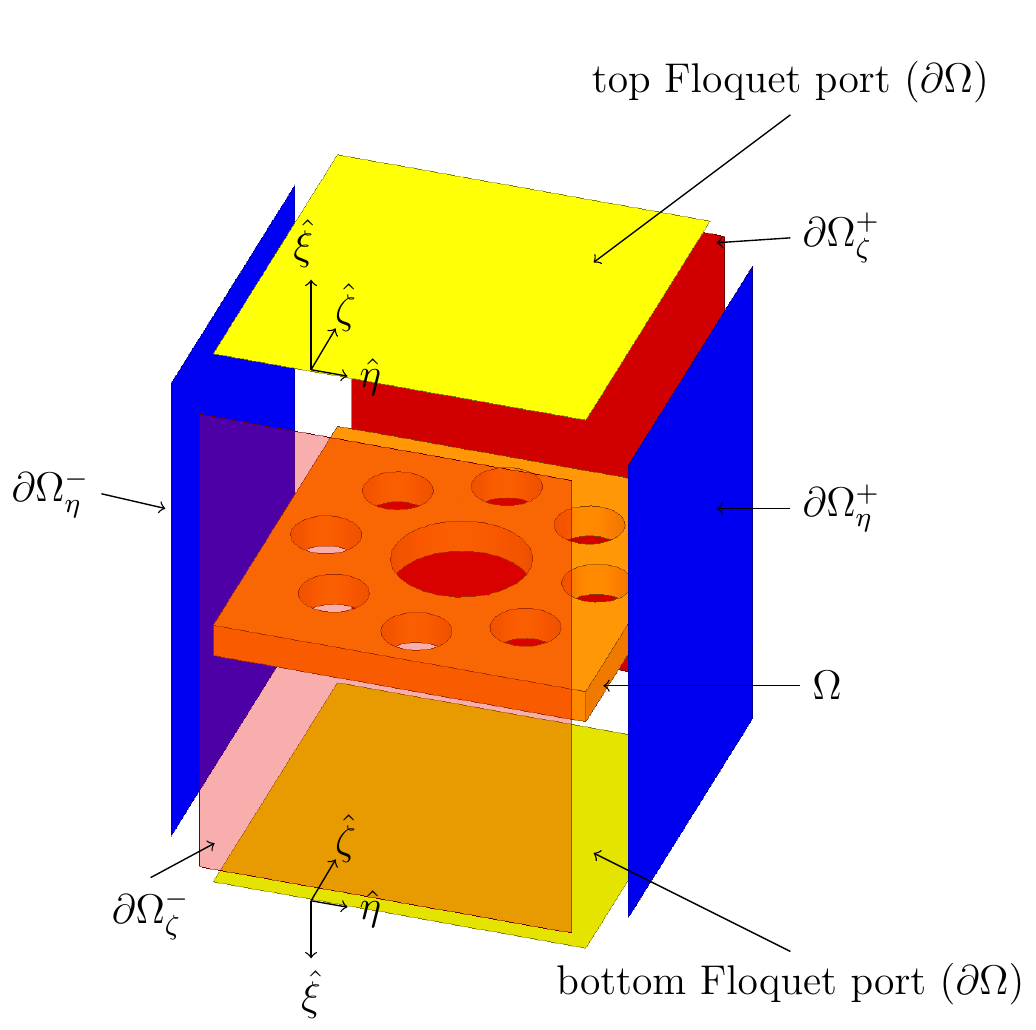}}
\caption{(a) Sample doubly-periodic structure under consideration. Dashed blue box shows a single unit cell of the structure. (b) Unit cell of the periodic structure enclosed by a box with Floquet ports. Note that the box surfaces have been exploded for better visualization.}
\label{fig:Prd_illus}
\end{figure*}

Consider a doubly periodic system with a unit cell that occupies the volume $\Omega$ (Fig.~\ref{fig:Prd_illus}). Let $\partial \Omega$ denote the interfaces of $\Omega$ to the medium surrounding the system. Once again, assume the vicinity of $\partial \Omega$ is parameterized by a locally Cartesian coordinate system $(\xi,\eta,\zeta)$ with $\zeta > 0$ exterior to $\Omega$.  Assume that $\Omega$ measures $L_\eta \times L_\zeta \times L_\xi$ and let $\partial \Omega_\eta^{\pm}$ and $\partial \Omega_\zeta^{\pm}$ denote the interfaces of $\Omega$ to neighboring cells. Finally, suppose that $\Omega$ is filled with a nondispersive and lossless material with permittivity $\varepsilon(\vr)$ and permeability $\mu(\vr)$, both of which are constant near $\partial \Omega$. Fields near $\partial \Omega$ can be expressed in terms of $M$ Floquet modes with frequency independent and complex-valued mode profiles $\bm{\mathcal{X}}_p$ (Appendix \ref{Appdix:Floquet_mode}), $\xi-$propagation constants $\beta_p$, and $\xi-$impedances $Z_p$, $p=1, ..., M$.  Just as in the preceding section, some of these modes may be evanescent. The mode profiles satisfy the orthonormality condition 
\begin{align}
\int_{\partial \Omega} \bm{\mathcal{X}}_p^{*}(\eta,\zeta) \cdot \bm{\mathcal{X}}_q (\eta,\zeta) d\eta d\zeta = \delta_{pq} \,. \label{eq:Floquet_orthogonal}
\end{align}
(Note the conjugation on $\bm{\mathcal{X}}_p$ in Eq.~\eqref{eq:Floquet_orthogonal} relative to Eq.~\eqref{eq:Guiding_orthogonal}).

With this setup, the treatment of the previous section can be repeated wholesale, provided that the inter-cell boundaries $\partial \Omega_\eta^{\pm}$ and $\partial \Omega_\zeta^{\pm}$ are properly accounted for. Specifically, The transverse components of incoming, outgoing, and total fields still can be expanded in terms of Floquet modes as in Eqs.~\eqref{eq:incoming_E_transv} to \eqref{eq:tot_H_transv}, and manipulating Maxwell's equations still produces Eq.~\eqref{eq:Helmholtz_pre_integ}. That said, integrating the LHS and RHS of Eq.~\eqref{eq:Helmholtz_pre_integ} over $\Omega$ and applying the divergence theorem now yields
\begin{align}
&\frac{j}{2} \int_{\partial \Omega_\eta^{-}} (-\hat{\eta}) \cdot \left( \fE_{p}' \times \fH_{q}^* + \fE_{q}^* \times \fH_{p}' \right) dS \nonumber \\
& \quad + \frac{j}{2} \int_{\partial \Omega_\eta^{+}} \hat{\eta} \cdot \left( \fE_{p}' \times \fH_{q}^* + \fE_{q}^* \times \fH_{p}' \right) dS \nonumber \\
& \quad + \frac{j}{2} \int_{\partial \Omega_\zeta^{-}} (-\hat{\zeta}) \cdot \left( \fE_{p}' \times \fH_{q}^* + \fE_{q}^* \times \fH_{p}' \right) dS \nonumber \\
& \quad + \frac{j}{2} \int_{\partial \Omega_\zeta^{+}} \hat{\zeta} \cdot \left( \fE_{p}' \times \fH_{q}^* + \fE_{q}^* \times \fH_{p}' \right) dS \nonumber \\
& \quad + \frac{j}{2} \int_{\partial \Omega} \hat{\xi} \cdot \left( \fE_{p}' \times \fH_{q}^* + \fE_{q}^* \times \fH_{p}' \right) dS \nonumber \\
& \quad \quad \quad = \frac{1}{2} \int_{\Omega} \left( \varepsilon \fE_q^* \cdot \fE_p + \mu \fH_q^* \cdot \fH_p \right) dV \,. \label{eq:Helmholtz_B_integ1}
\end{align}

To evaluate the LHS of Eq. (23), note that the total fields on boundaries $\partial \Omega_\eta^\pm$ and $\partial \Omega_\zeta^\pm$ satisfy the periodic boundary conditions, \ie $\fE_{p,q} |_{\partial \Omega_\eta^+} = \fE_{p,q} |_{\partial \Omega_\eta^-}$ and $\fE_{p,q} |_{\partial \Omega_\zeta^+} = \fE_{p,q} |_{\partial \Omega_\zeta^-}$. Similar conditions hold for the magnetic fields.
It is clear that $\left( \fE_{p}' \times \fH_{q}^* + \fE_{q}^* \times \fH_{p}' \right) \Bigr|_{\partial \Omega_\eta^{+}} = \left( \fE_{p}' \times \fH_{q}^* + \fE_{q}^* \times \fH_{p}' \right) \Bigr|_{\partial \Omega_\eta^{-}}$ and this also holds for the boundaries $\partial \Omega_\zeta^{\pm}$.  

It follows that the surface integrals over the periodic boundaries $\partial \Omega_\eta^\pm$, $\partial \Omega_\zeta^\pm$ in Eq.~\eqref{eq:Helmholtz_B_integ1} cancel, and that only those over the Floquet ports $\partial \Omega$ remains. Eq.~\eqref{eq:Helmholtz_B_integ1} hence reverts back to Eq.~\eqref{eq:Helmholtz_g_integ2} and the remainder of the derivation for guiding systems again carries over to periodic ones. The generalized WS relationships Eqs.~\eqref{eq:WS_gen2} and \eqref{eq:volumeQ_plus_correction}, derived earlier for guiding systems with PEC walls, therefore also apply to periodic systems as well.

\subsection{Computing $\matr{S}'$ from $\matr{S}$ and $\matr{Q}$ without Matrix Inversion}
\label{sec:em_dw_S}

In the absence of evanescent modes, the scattering matrix $\matr{S}$ is unitary ($\matr{S}^\dag \matr{S} = \matr{I}$) and Eq.~\eqref{eq:Q_def1} can be used to characterize $\matr{S}$'s frequency sensitivity without matrix inversion as $\matr{S}' = j \matr{S} \matr{Q}$. This section shows that in the presence of evanescent modes, Eq.~\eqref{eq:WS_gen2} still allows for the computation of $\matr{S}'$ without relying on matrix inversion. 

To demonstrate this useful factoid, let $\matr{\Gamma}$ denote
\begin{align}
\matr{\Gamma} = \left( \matr{N}_{-} +  \matr{S}^\dag \matr{N}_{+} \right)^{-1} \,. \label{eq:Gamma}
\end{align}
For notational convenience, the indices of the propagating modes $\text{(P)}$ are assumed smaller than those of the evanescent ones $\text{(E)}$, and matrices $\matr{S}$, $\matr{N}_{-}$, $\matr{S}^\dag \matr{N}_{+}$ and $\matr{\Gamma}$ are partitioned as 
\begin{align}
\matr{S} = 
\left(
\begin{array}{cc}
\matr{S}_\text{PP} & \matr{S}_\text{PE} \\
\matr{S}_\text{EP} & \matr{S}_\text{EE}
\end{array}
\right) \label{eq:S_blocks}
\end{align}
\begin{align}
\matr{N}_{-}
&=
\left(
\begin{array}{cc}
\matr{0}_\text{PP} & \\
 & \matr{I}_\text{EE}
\end{array}
\right) \label{eq:N_minus}
\end{align}
\begin{align}
\matr{S}^\dag \matr{N}_{+}
&=
\left(
\begin{array}{cc}
j \matr{S}_\text{PP}^\dag & \matr{0}_\text{PE} \\
j \matr{S}_\text{PE}^\dag & \matr{0}_\text{EE}
\end{array}
\right) \label{eq:S_dag_N_plus}
\end{align}
\begin{align}
\matr{\Gamma}
&=
\left(
\begin{array}{cc}
\matr{\Gamma}_\text{PP} & \matr{\Gamma}_\text{PE} \\
\matr{\Gamma}_\text{EP} & \matr{\Gamma}_\text{EE}
\end{array}
\right) \,.
\end{align}
Adding Eqs.~\eqref{eq:N_minus} and \eqref{eq:S_dag_N_plus}, and multiplying the result by Eq.~\eqref{eq:Gamma} produces
\begin{align}
\left(
\begin{array}{cc}
\matr{\Gamma}_\text{PP} & \matr{\Gamma}_\text{PE} \\
\matr{\Gamma}_\text{EP} & \matr{\Gamma}_\text{EE}
\end{array}
\right)
\left(
\begin{array}{cc}
j \matr{S}_\text{PP}^\dag & \\
j \matr{S}_\text{PE}^\dag & \matr{I}_\text{EE}
\end{array}
\right)
=
\left(
\begin{array}{cc}
\matr{I}_\text{PP} & \\
 & \matr{I}_\text{EE}
\end{array}
\right) \,. \label{eq:Gamma_condition}
\end{align}
Finally, solving Eq.~\eqref{eq:Gamma_condition} for $\matr{\Gamma}$ yields
\begin{align}
\matr{\Gamma} = 
\left(
\begin{array}{cc}
-j \matr{S}_\text{PP} & \matr{0} \\
- \matr{S}_\text{PE}^\dag \matr{S}_\text{PP} & \matr{I}_\text{EE}
\end{array}
\right) \, \label{eq:matrix_Gamma}
\end{align}
where use was made of the fact that $\matr{S}_\text{PP}$ is unitary ($\matr{S}_\text{PP}^\dag \matr{S}_\text{PP} = \matr{I}_\text{PP}$) in accordance with the lossless nature of the system. It immediately follows that $\matr{S}'$ can be computed via
\begin{align}
\matr{S}' = \matr{\Gamma} \matr{Q} \label{eq:WS_Floquet_dw_dS}
\end{align}
without performing any matrix inversion.

\subsection{Time Delay Interpretation of the Generalized WS Relationship and WS Modes}
\label{sec:em_interpret}

In the abscence of evanescent modes, $\matr{Q}$ can be diagonalized as $\matr{Q} = \matr{W} \overline{\matr{Q}} \matr{W}^\dag$ where the columns of $\matr{W}$ and the diagonal elements of $\overline{\matr{Q}}$ describe so-called WS modes and their time delays, viz. group delays experienced by narrowband wave packets that interact with the system \cite{Patel2020WS1}. In the presence of evanescent modes, this property of $\matr{Q}$, now defined via Eq.~\eqref{eq:WS_gen2}, needs to be re-examined. This section shows that the generalized WS time delay matrix still allows for the construction of modes with well defined time delays by diagonalizing the block of the WS time delay matrix that accounts for interactions between propagating modes, properly modified to account for the presence of evanescent ones. 

To demonstrate this concept, Eq.~\eqref{eq:WS_gen2} is expressed in block form, leveraging notation introduced in the preceding section:
\begin{align}
\matr{Q}_\text{prop} &\equiv \widetilde{\matr{Q}}_\text{PP} + \matr{S}_\text{PP}^\dag \matr{\Lambda}_\text{PP} - \matr{\Lambda}_\text{PP} \matr{S}_\text{PP} - \matr{S}_\text{EP}^\dag \matr{\Lambda}_\text{EE} \matr{S}_\text{EP} = j \matr{S}_\text{PP}^\dag \matr{S}'_\text{PP} \label{eq:Q_prop} \\
& \widetilde{\matr{Q}}_\text{PE} - \matr{\Lambda}_\text{PP} \matr{S}_\text{PE} - \matr{S}_\text{EP}^\dag \matr{\Lambda}_\text{EE} \matr{S}_\text{EE} = j \matr{S}_\text{PP}^\dag \matr{S}'_\text{PE} \label{eq:Q12} \\
& \widetilde{\matr{Q}}_\text{EP} + \matr{S}_\text{PE}^\dag \matr{\Lambda}_\text{PP} - \matr{S}_\text{EE}^\dag \matr{\Lambda}_\text{EE} \matr{S}_\text{EP} = j \matr{S}_\text{PE}^\dag \matr{S}'_\text{PP} + \matr{S}'_\text{EP} \label{eq:Q21} \\
& \quad \widetilde{\matr{Q}}_\text{EE} + \matr{\Lambda}_\text{EE} - \matr{S}_\text{EE}^\dag \matr{\Lambda}_\text{EE} \matr{S}_\text{EE} = j \matr{S}_\text{PE}^\dag \matr{S}'_\text{PE} + \matr{S}'_\text{EE} \,. \label{eq:Q22}
\end{align}
Here $\matr{\Lambda}_\text{PP} = \left(\matr{\Lambda}_{-}\right)_\text{PP}$ and $\matr{\Lambda}_\text{EE} = \left(\matr{\Lambda}_{+}\right)_\text{EE}$. 
Eq.~\eqref{eq:Q_prop} describes how $\widetilde{\matr{Q}}_\text{PP}$, the submatrix of $\widetilde{\matr{Q}}$ corresponding to only propagating modes, relates to submatrices $\matr{S}_\text{PP}$ and $\matr{S}_\text{PP}'$. 
Compared to the exposition in Eq.~\eqref{eq:Q_prop} \cite{Patel2020WS1}, $\matr{Q}_\text{prop}$ now includes an additional term $\matr{S}_\text{EP}^\dag \matr{\Lambda}_\text{EE} \matr{S}_\text{EP}$ that accounts for interactions between propagating and evanescent modes.

The matrices $\matr{Q}_\text{prop}$ and $\matr{S}_\text{PP}$ in satisfy three important properties: 
\begin{enumerate}
    \item $\matr{Q}_\text{prop}$ is Hermitian; this fact follows directly from its defining Eq.~\eqref{eq:Q_prop};
    \item $\matr{S}_\text{PP}$ is unitary due to the lossless nature of the system;
    \item $\matr{S}_\text{PP}$ is symmetric due to the reciprocal nature of the system.
\end{enumerate}
These properties allow the relationship $\matr{Q}_\text{prop} = j \matr{S}_\text{PP}^\dag \matr{S}'_\text{PP} $ to be interpreted using the techniques presented in \cite{Patel2020WS1}.  Specifically, it follows that the matrices $\matr{Q}_\text{prop}$, $\matr{S}_\text{PP}$ and $\matr{S}_\text{PP}'$ can be simultaneously diagonalized using $\matr{Q}_\text{prop}$'s eigenvector matrix $\matr{W}$ as 
\begin{align}
\matr{Q}_\text{prop} &= \matr{W} \overline{\matr{D}} \matr{W}^\dag \label{eq:Q_prop_diagonalize} \\
\matr{S}_\text{PP} &= \matr{W}^* \overline{\matr{S}_\text{PP}} \matr{W}^\dag \label{eq:S_pp_diagonalize} \\
\matr{S}_\text{PP}' &= \matr{W}^* \overline{\matr{S}_\text{PP}'} \matr{W}^\dag \label{eq:dw_S_pp_diagonalize}
\end{align}
where $\overline{\matr{D}}$, $\overline{\matr{S}_\text{PP}}$ and $\overline{\matr{S}_\text{PP}'}$ are diagonal matrices. $\overline{\matr{D}}$ comprises $\matr{Q}_\text{prop}$'s eigenvalues, which can be interpreted as time/group delays experienced by WS modes obtained by weighing waveguide modes by the entries of the columns of $\matr{W}$. The WS modes are fully decoupled and exhibit minimal dispersion upon interacting with the system \cite{Patel2020WS1}.

For periodic systems, the mode profiles are complex, the orthogonality condition \eqref{eq:Floquet_orthogonal} involves conjugation, and $\matr{S}_\text{PP}$ becomes asymmetric. That said, it can be trivially shown that $\matr{S}_\text{PP}$ becomes symmetric upon row permutation. Specifically, the matrix $\matr{I}_r \matr{S}_\text{PP}$ where  
\begin{align}
\left( \matr{I}_r \right)_{i_p, i_q} = \delta_{p \bar{q}} \,.
\end{align}
where $i_p$ and $i_q$ are the indices of modes $p$ and $q$ in $\matr{S}_\text{PP}$, and $\bar{q}$ is the index that satisfies $\bm{\mathcal{X}}_q = \bm{\mathcal{X}}^{*}_{\bar{q}}$, is symmetric. Upon re-expressing Eq.~\eqref{eq:Q_prop} as $\matr{Q}_\text{prop} = j \left( \matr{I}_r \matr{S}_\text{PP} \right)^\dag \left( \matr{I}_r \matr{S}_\text{PP}' \right)$, it follows that the simultaneous diagonalization in  Eqs.~\eqref{eq:Q_prop_diagonalize}--\eqref{eq:dw_S_pp_diagonalize} can proceed upon replacing $\matr{S}_\text{PP}$ and $\matr{S}_\text{PP}'$ by $\matr{I}_r \matr{S}_\text{PP}$ and $\matr{I}_r \matr{S}_\text{PP}'$, respectively.

Finally, note that whereas $\matr{Q}_\text{prop}$ allows for the construction of WS with well defined time delays, $\matr{Q}$'s other submatrices $\matr{Q}_\text{PE}$, $\matr{Q}_\text{EP}$ and $\matr{Q}_\text{EE}$ appear to have no direct physical interpretation.

\subsection{Evaluating the WS Time Delay Matrix of Compound Systems From Those of Its Subsystems}
\label{sec:cascade}

This section demonstrates that the WS time delay matrix of a composite system can be easily obtained from those of its subsystems. 

Consider a composite system obtained by cascading several subsystems through interfaces that support evanescent fields. This section demonstrates that the WS time delay matrix of the composite systems can be obtained from those of the subsystems. Only the case of two subsystems is considered as additional subsystems can be accounted for through recursion. 

Eq.~\eqref{eq:volumeQ_plus_correction} shows that the WS time delay matrix $\matr{Q}$ consists of a computationally expensive volume-integral component $\widetilde{\matr{Q}}$ as well as easily computed correction terms that relate to port impedances and their corresponding elements in the scattering matrix. Our objective therefore is to compute $\widetilde{\matr{Q}}$ of the composite system from those of its subsystems, since the correction terms are trivially accounted for.  

\begin{figure}[htbp!]
\centering\includegraphics[width=7.0cm]{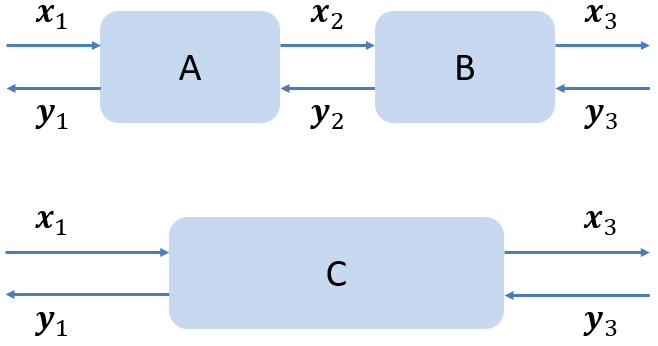}
\caption{Two smaller systems, \textit{A} and \textit{B}, are cascaded to form a bigger system C. $\bm{x}_i$ and $\bm{y}_i$ are combination coefficient vectors of ``forward-going'' and ``backward-going'' waves, respectively.}
\label{fig:cascade_illus}
\end{figure}

Consider a system \textit{C} obtained by cascading two subsystems \textit{A} and \textit{B}. In what follows, the mode indices of systems \textit{A} and \textit{B} are arranged and their scattering matrices partitioned in accordance with the labeling of waves in Fig.~\ref{fig:cascade_illus}, as
\begin{align}
\matr{S}^A &=
\left(
\begin{array}{cc}
\matr{S}^A_{11} & \matr{S}^A_{12} \\
\matr{S}^A_{21} & \matr{S}^A_{22}
\end{array} \right) \,,
\end{align}
\begin{align}
\mathbf{S}^B &= 
\left(
\begin{array}{cc}
\matr{S}^B_{22} & \matr{S}^B_{23} \\
\matr{S}^B_{32} & \matr{S}^B_{33}
\end{array} \right)
\end{align} 
and
\begin{align}
\mathbf{S}^C &= 
\left(
\begin{array}{cc}
\matr{S}^C_{11} & \matr{S}^C_{13} \\
\matr{S}^C_{31} & \matr{S}^C_{33}
\end{array} \right) \,.
\end{align} 
Similar decompositions exist for $\widetilde{\matr{Q}}^A$, $\widetilde{\matr{Q}}^B$, $\widetilde{\matr{Q}}^C$.

Using methods for computing $\matr{S}^C$ from $\matr{S}^A$ and $\matr{S}^B$ described in \cite{Chu1986generalized, Neilson1989Deter}, it is easy to show that the amplitudes of waves traveling in between systems \textit{A} and \textit{B} can be expressed as
\begin{align}
\mathbf{x}_2 &= \matr{T}_x \mathbf{x}_1 + \matr{R}_y \mathbf{y}_3 \label{eq:x2_x1_y3} \\
\mathbf{y}_2 &= \matr{T}_y \mathbf{y}_3 + \matr{R}_x \mathbf{x}_1
\label{eq:y2_y3_x1}
\end{align}
where
\begin{align}
\matr{R}_x &= (\matr{I} - \matr{S}^B_{22} \matr{S}^A_{22})^{-1} \matr{S}^B_{22} \matr{S}^A_{21} \\
\matr{T}_x &= (\matr{I} - \matr{S}^A_{22} \matr{S}^B_{22})^{-1} \matr{S}^A_{21} \\
\matr{R}_y &= (\matr{I} - \matr{S}^A_{22} \matr{S}^B_{22})^{-1} \matr{S}^A_{22} \matr{S}^B_{23} \\
\matr{T}_y &= (\matr{I} - \matr{S}^B_{22} \matr{S}^A_{22})^{-1} \matr{S}^B_{23} \,.
\end{align}

Let $\fE_p^\alpha$ denote the electric field in system $\alpha$  ($\alpha = A,B,C$) when excited by mode $p$ ($p \in \{ A1, A2, B2, B3\}$). It follows from Eqs.~\eqref{eq:x2_x1_y3} and \eqref{eq:y2_y3_x1} that the electric field in composite system $C$ can be expressed in terms of those in subsystems $A$ and $B$ as
\begin{align}
\fE_p^C &= 
\fE_p^A + \sum_{m \in A2} \fE_m^A \left(\mathbf{R}_x\right)_{mp} + \sum_{m \in B2} \fE_m^B \left(\mathbf{T}_x\right)_{mp} \label{eq:EP_C_A1}
\end{align}
for $p \in A1$, or
\begin{align}
\fE_p^C &= 
\fE_p^B + \sum_{m \in B2} \fE_m^B \left(\mathbf{R}_y\right)_{mp} + \sum_{m \in A2} \fE_m^A \left(\mathbf{T}_y\right)_{mp} \label{eq:EP_C_B3}
\end{align}
for $p \in B3$. 
Here $\left(\mathbf{R}_{x,y}\right)_{mp}$ and $\left(\mathbf{T}_{x,y}\right)_{mp}$ denote matrix elements corresponding to the $m$-th and $p$-th modes, instead of the $m$-th row and $p$-th column. Magnetic fields in composite system $C$ can be obtained similarly.

Using Eq.~\eqref{eq:Q_tilde} with respect to the composite system $C$ and substituting Eqs.~\eqref{eq:EP_C_A1}--\eqref{eq:EP_C_B3} yields
\begin{align}
\widetilde{Q}_{qp}^C 
& = \widetilde{Q}_{qp}^A + \sum_{m \in A2} \widetilde{Q}_{qm}^A \left(\mathbf{R}_x\right)_{mp}
+ \sum_{m \in A2} \left(\mathbf{R}_x^\dag\right)_{mq} \widetilde{Q}_{mp}^A \nonumber \\
& \quad + \sum_{m \in A2} \sum_{n \in A2} \left(\mathbf{R}_x^\dag\right)_{mq} \widetilde{Q}_{mn}^A \left(\mathbf{R}_x\right)_{np} \nonumber \\
& \quad + \sum_{m \in B2} \sum_{n \in B2} \left(\mathbf{T}_x^\dag\right)_{mq} \widetilde{Q}_{mn}^B \left(\mathbf{T}_x\right)_{np} \label{eq:QC_A1A1}
\end{align}
for $q \in A1$ and $p \in A1$,
\begin{align}
\widetilde{Q}_{qp}^C 
&= \sum_{m \in A2} \left(\mathbf{T}_y^\dag\right)_{mq} \widetilde{Q}_{mp}^A
+ \sum_{m \in A2} \sum_{n \in A2} \left(\mathbf{T}_y^\dag\right)_{mq} \widetilde{Q}_{mn}^A \left(\mathbf{R}_x\right)_{np} \nonumber \\
& + \sum_{m \in B2} \sum_{n \in B2} \left(\mathbf{R}_y^\dag\right)_{mq} \widetilde{Q}_{mn}^B \left(\mathbf{T}_x\right)_{np}
+ \sum_{m \in B2} \widetilde{Q}_{qm}^B \left(\mathbf{T}_x\right)_{mp} \label{eq:QC_A1B3}
\end{align} 
for $q \in A1$ and $p \in B3$,
\begin{align}
\widetilde{Q}_{qp}^C 
&= \sum_{m \in A2} \widetilde{Q}_{qm}^A \left(\mathbf{T}_y\right)_{mp}
+ \sum_{m \in A2} \sum_{n \in A2} \left(\mathbf{R}_x^\dag\right)_{mq} \widetilde{Q}_{mn}^A \left(\mathbf{T}_y\right)_{np} \nonumber \\
& + \sum_{m \in B2} \sum_{n \in B2} \left(\mathbf{T}_x^\dag\right)_{mq} \widetilde{Q}_{mn}^B \left(\mathbf{R}_y\right)_{np}
+ \sum_{m \in B2} \left(\mathbf{T}_x^\dag\right)_{mq} \widetilde{Q}_{mp}^B \label{eq:QC_B3A1}
\end{align}
for $q \in B3$ and $p \in A1$, and 
\begin{align}
\widetilde{Q}_{qp}^C 
&= \sum_{m \in A2} \sum_{n \in A2} \left(\mathbf{T}_y^\dag\right)_{mq} \widetilde{Q}_{mn}^A \left(\mathbf{T}_y\right)_{np} \nonumber \\
& \quad + \sum_{m \in B2} \sum_{n \in B2} \left(\mathbf{R}_y^\dag\right)_{mq} \widetilde{Q}_{mn}^B \left(\mathbf{R}_y\right)_{np} \nonumber \\
& \quad + \sum_{m \in B2} \left(\mathbf{R}_y^\dag\right)_{mq} \widetilde{Q}_{mp}^B
+ \sum_{m \in B2} \widetilde{Q}_{qm}^B \left(\mathbf{R}_y\right)_{mp}
+ \widetilde{Q}_{qp}^B \label{eq:QC_B3B3}
\end{align}
for $q \in B3$ and $p \in B3$. 
Eqs.~\eqref{eq:QC_A1A1}--\eqref{eq:QC_B3B3} can be compactly expressed as
\begin{align}
\widetilde{\mathbf{Q}}^C 
&= \mathbf{P}_A^\dag \widetilde{\mathbf{Q}}^A \mathbf{P}_A + \mathbf{P}_B^\dag \widetilde{\mathbf{Q}}^B \mathbf{P}_B \label{eq:Q_cscd}
\end{align}
where
\begin{align}
\mathbf{P}_A &=
\left(
\begin{array}{cc}
\mathbf{I} & \mathbf{0} \\
\mathbf{R}_x & \mathbf{T}_y
\end{array} \right) \\
\mathbf{P}_B &= 
\left(
\begin{array}{cc}
\mathbf{T}_x & \mathbf{R}_y \\
\mathbf{0} & \mathbf{I}
\end{array} \right) \,. \label{eq:Q_cscd_P_B}
\end{align}

Eqs.~\eqref{eq:Q_cscd}--\eqref{eq:Q_cscd_P_B} constitute the sought equations for obtaining the WS time delay matrix of a composite system from those of its subsystems.

\section{Numerical Examples}
\label{sec:example}

This section includes three numerical examples that validates the WS techniques in Section \ref{sec:em}. First the generalized WS relationships are verified for a waveguide containing a scatterer within close proximity of its port, necessitating the incorporation of evanescent modes into the scattering matrix. Next, the cascade formulation is validated via its application to two coupled cavities. Finally, the generalized WS relationships are verified for a periodic photonic crystal slab containing through-waveguides. All guiding systems are simulated using an FEM leveraging mode expansions \cite{Jin2015finite}, while periodic systems are simulated by a finite-element boundary-integral (FEBI) method using periodic Green's functions  \cite{Eibert1999hybrid}.

\subsection{PEC-Terminated Waveguide with Object Near Its Port}

\begin{figure}[htbp!]
\centering\includegraphics[width=7.0cm]{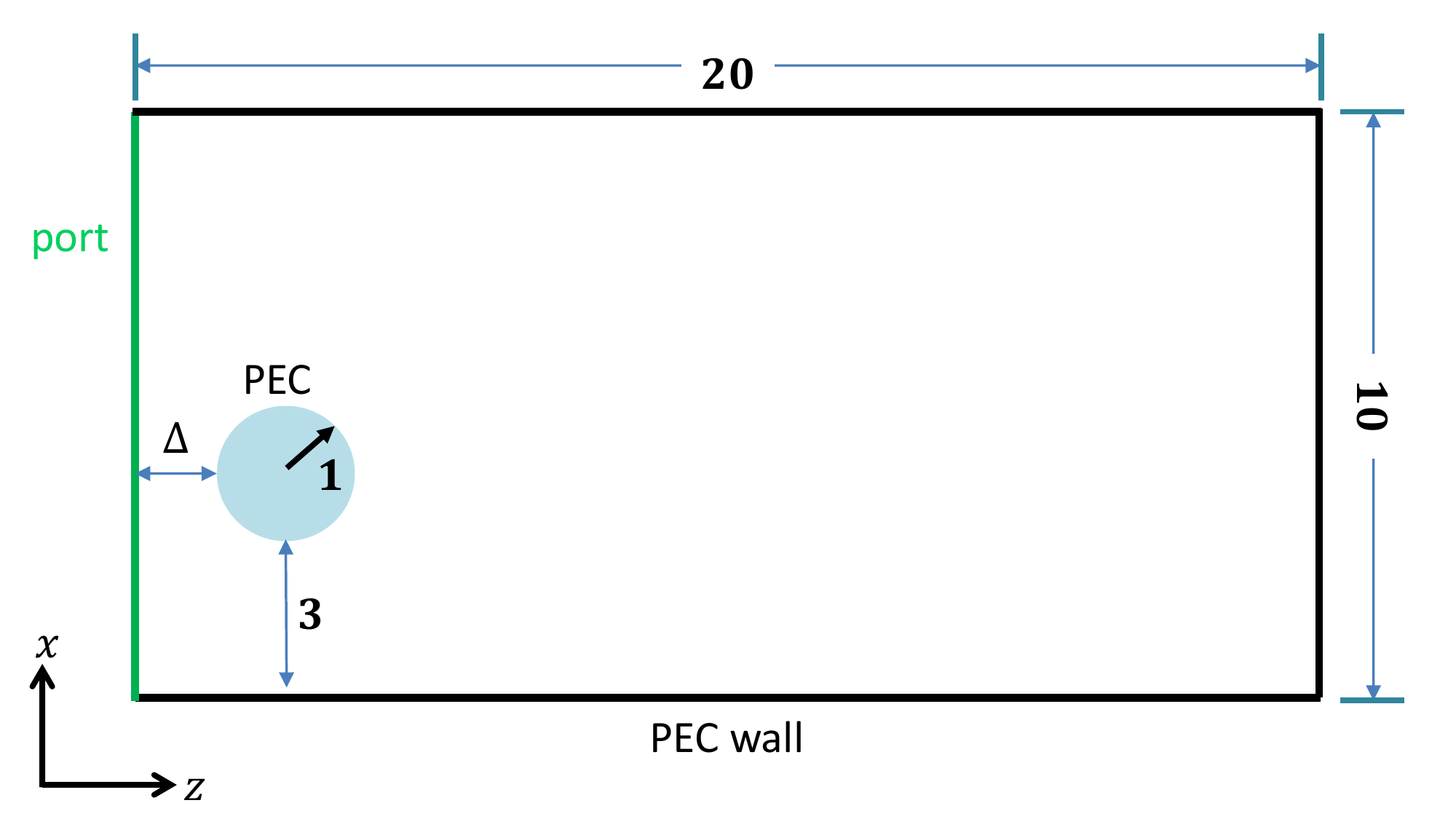}
\caption{A waveguide with a PEC object near the port (unit: mm). All materials are homogeneous in the $y$-direction.}
\label{fig:case1_circ}
\end{figure}

Consider the air-filled and PEC-terminated waveguide shown in Fig.~\ref{fig:case1_circ} that contains a PEC cylinder located a small distance $\Delta$ from its port. At $f=1.428 \times 10^{11}$ Hz the waveguide supports 9 propagating TE modes.

\begin{figure}[htbp!]
\centering\includegraphics[width=7.0cm]{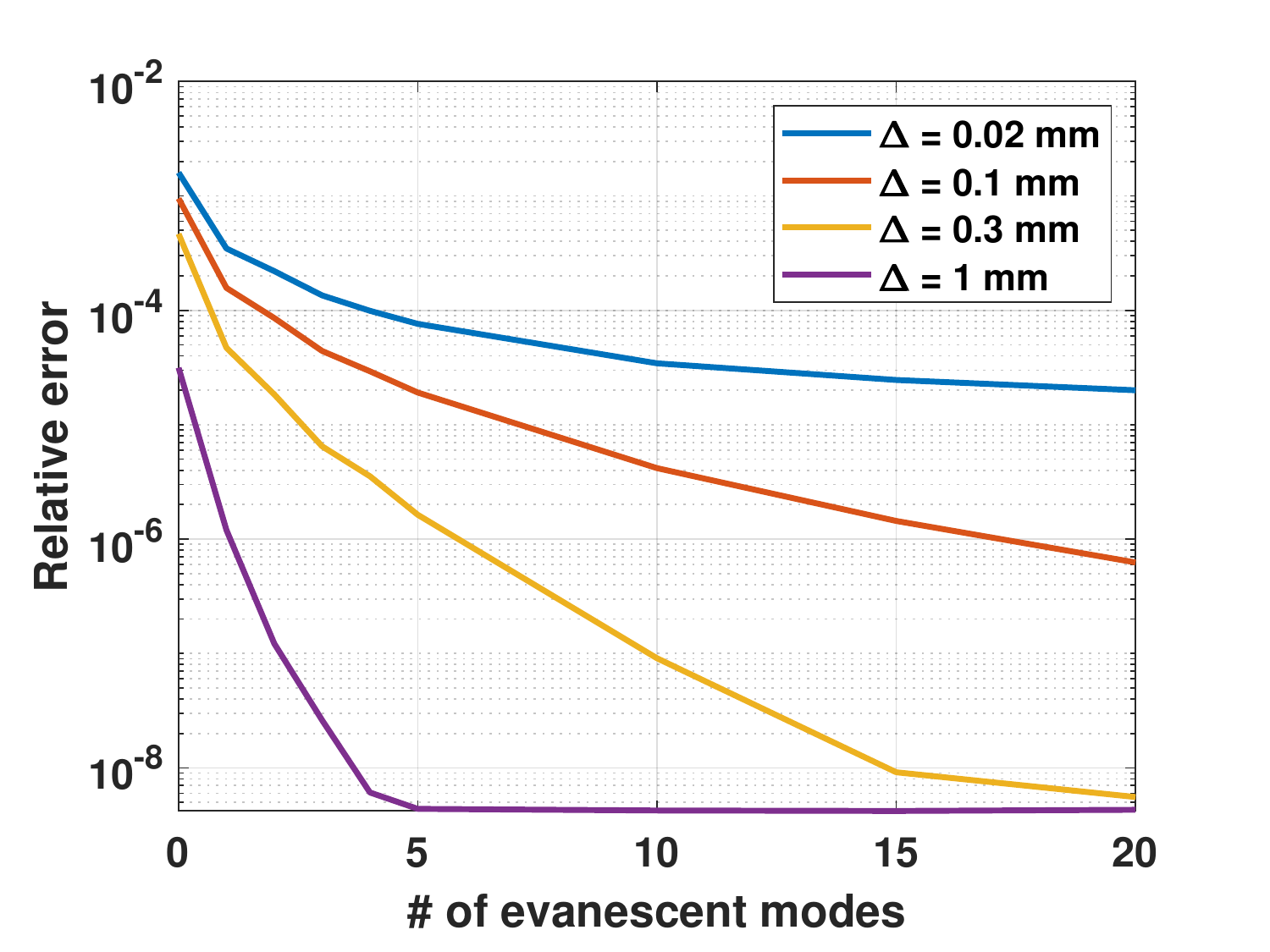}
\caption{Relative error in WS relationship Eq.~\eqref{eq:WS_gen2}, $\| \matr{Q} - (\matr{N}_{-} + \matr{S}^\dag \matr{N}_{+}) \matr{S}'  \|_F / \| \matr{Q} \|_F$ versus the number of evanescent modes $M-9$ for various $\Delta$.}
\label{fig:case1_convg_circ}
\end{figure}

First, the accuracy of generalized WS relationship \eqref{eq:WS_gen2} is verified for different choices of $\Delta$ and $M-9$, the number of evanescent modes. Fig.~\ref{fig:case1_convg_circ} shows that the accuracy of the WS relationship improves as $\Delta$ and/or $M-9$ increase.

\begin{figure}[htbp!]
\centering\includegraphics[width=6.5cm]{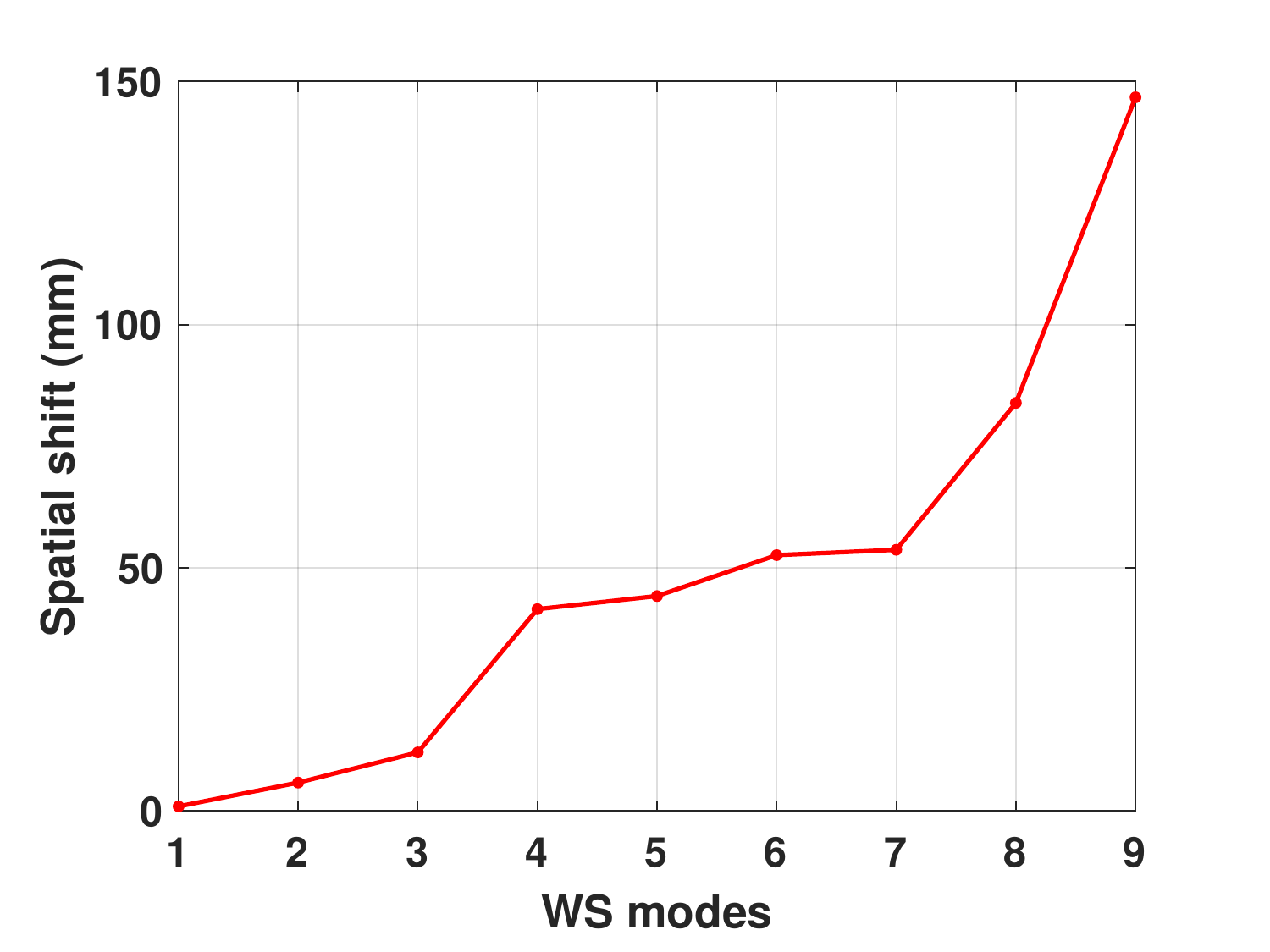}
\caption{Spatial shifts $\text{diag}(c \overline{\matr{D}})$ converted from time delays by multiplication with free space light speed, for the PEC-terminated waveguide with an object near the port.}
\label{fig:case1_time_delay}
\end{figure}

\begin{figure}[htbp!]
\null \hfill
\subfloat[Mode \#1, 0.9441 mm \label{fig:case1_WSmode_1}]{\includegraphics[width=0.5\columnwidth]{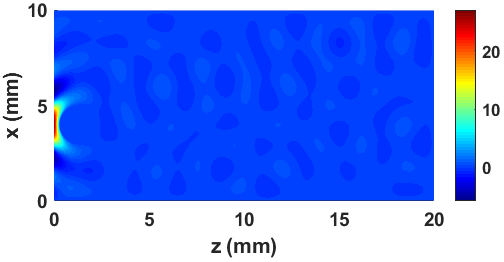}} \hfill
\subfloat[Mode \#4, 41.51 mm \label{fig:case1_WSmode_4}]{\includegraphics[width=0.5\columnwidth]{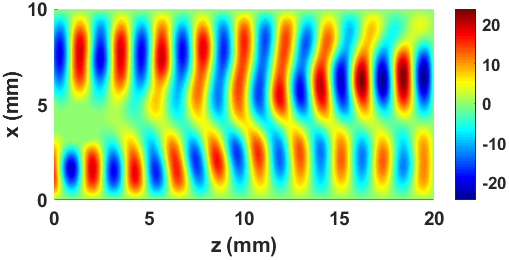}} \hfill
\hfill \null \\
\null \hfill
\subfloat[Mode \#6, 52.62 mm \label{fig:case1_WSmode_6}]{\includegraphics[width=0.5\columnwidth]{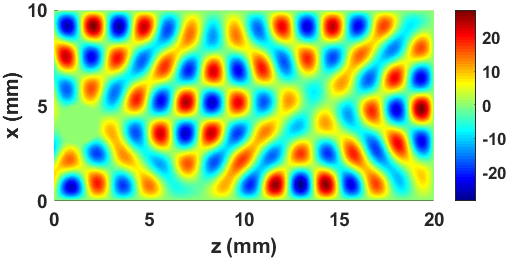}} \hfill
\subfloat[Mode \#9, 146.8 mm \label{fig:case1_WSmode_9}]{\includegraphics[width=0.5\columnwidth]{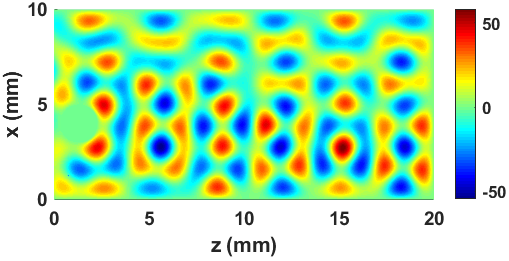}} \hfill
\caption{Several WS modes $real(E_y)$ of the PEC-terminated waveguide with an object near its port. Distance values below each subfigures are the corresponding equivalent spatial shifts in Fig.~\ref{fig:case1_time_delay}.}
\label{fig:case1_WSmodes}
\end{figure}

Next, assume $\Delta = 0.3$ mm and $M=24$. The relative error in $\matr{Q}$  (defined in the caption of Fig. 5) is on the order of $10^{-8}$. The $9 \times 9$ matrix $\matr{Q}_\text{prop}$ in Eq.~\eqref{eq:Q_prop} is diagonalized as $\matr{Q}_\text{prop} = \matr{W} \overline{\matr{D}} \matr{W}^\dag$; the columns of $\matr{W}$ and diagonal elements of $\overline{\matr{D}}$ describe WS modes and their time delays, respectively. Fig.~\ref{fig:case1_time_delay} shows the WS modes' ``spatial shifts'', obtained by multiplying time delays by the free-space speed of light. Fig.~\ref{fig:case1_WSmodes} depicts several WS modes. 

\begin{enumerate}
\item WS mode \#1 experiences the smallest time delay, corresponding to a spatial shift of roughly $2 \Delta$, as it describes a localized port excitation that exits the system after traveling to the cylinder and back (Fig.~\ref{fig:case1_WSmode_1}). 
\item WS mode \#4 experiences a time delay corresponding to a spatial shift that is slightly larger than two
round-trips between the port and the waveguide termination owing to the fact that it travels at a small angle w.r.t. the waveguide axis while avoiding interacting with the cylinder (Fig.~\ref{fig:case1_WSmode_4}). 
\item WS mode \#6 is very similar to mode \#4 except that it travels at a much larger angle w.r.t. the waveguide axis. The mode bounces not just off the waveguide termination, but also between its top and bottom walls, resulting in a much larger time delay. Interaction with the cylinder is avoided, however.
\item WS mode \#9 has the largest time delay because the wave not only reflects off the PEC walls but also interacts with the cylinder. It is trapped inside the waveguide for a significantly longer time than the other modes and exhibits semi-resonant behavior.  
\end{enumerate}

\subsection{Cascaded System}

\begin{figure}[htbp!]
\centering\includegraphics[width=8.0cm]{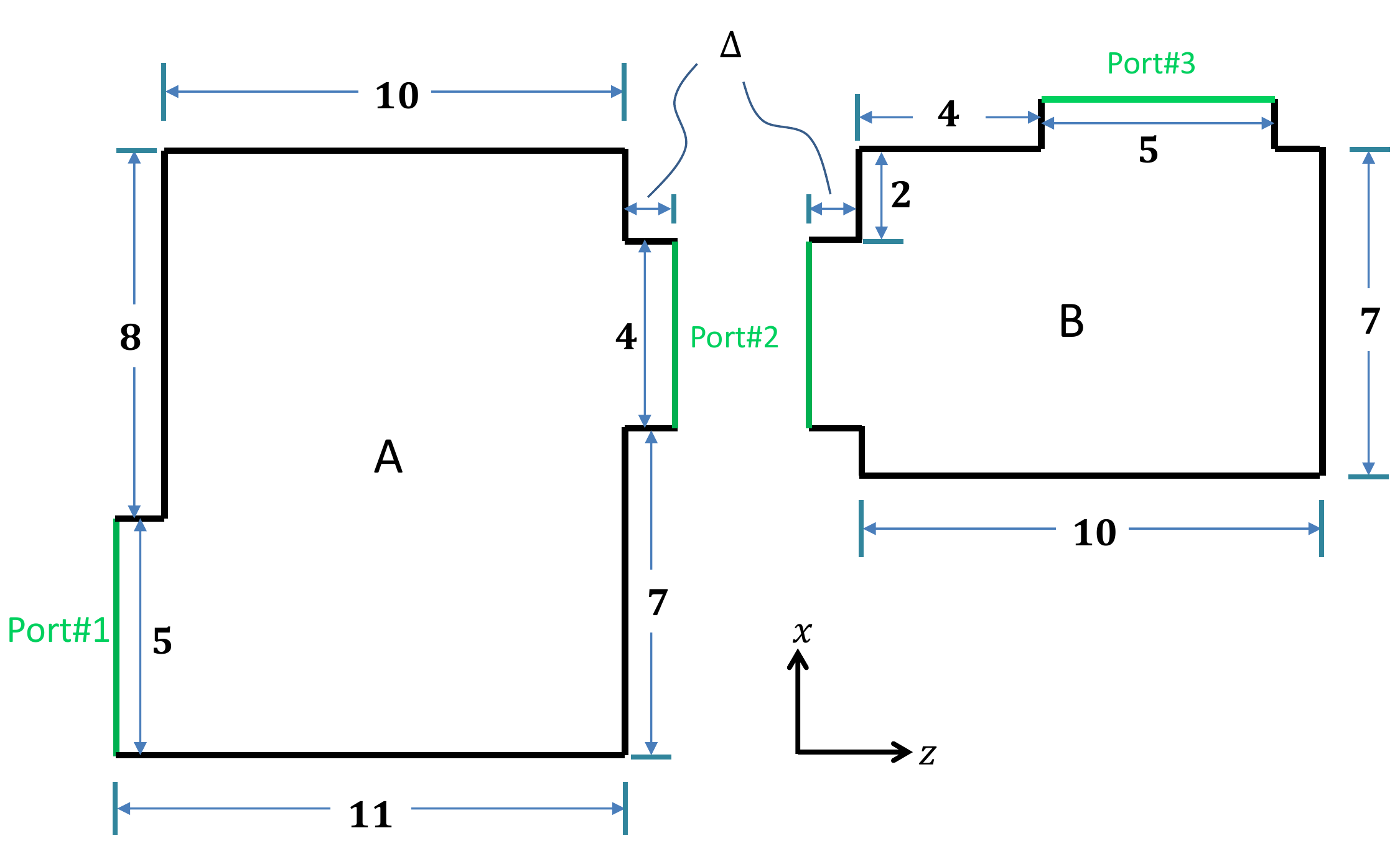}
\caption{Guiding system \textit{C} obtained by cascading (connecting) subsystems \textit{A} and \textit{B} at port \#2 (unit: mm).  All systems are air-filled and invariant along the $y$-direction, and $\Delta = 0.5$ mm.}
\label{fig:case2_AB}
\end{figure}

Consider the two air-filled guiding systems \textit{A} and \textit{B} shown in Fig.~\ref{fig:case2_AB}.  Guiding system \textit{C} is obtained by cascading systems \textit{A} and \textit{B} via port \#2. At $f=1.428 \times 10^{11}$ Hz, ports \#1 and \#3 both support 4 propagating TE modes while port \#2 supports 3 propagating TE modes.

$\matr{S}$ and $\matr{Q}$ for system \textit{C} are computed in two ways: (i) directly using the formulation of Section \ref{sec:em_guiding} resulting in $\matr{S}_C$ and $\matr{Q}_C$, and (ii) indirectly using the cascading procedure in Section \ref{sec:cascade} resulting in $\matr{S}_{A+B}$ and $\matr{Q}_{A+B}$. 
Fig. 9 shows that $\matr{S}_{A+B}$ and $\matr{Q}_{A+B}$ converge to $\matr{S}_C$ and $\matr{Q}_C$ as the number of evanescent modes accounted for in port \#2 increases.

\begin{figure}[htbp!]
\centering\includegraphics[width=7.0cm]{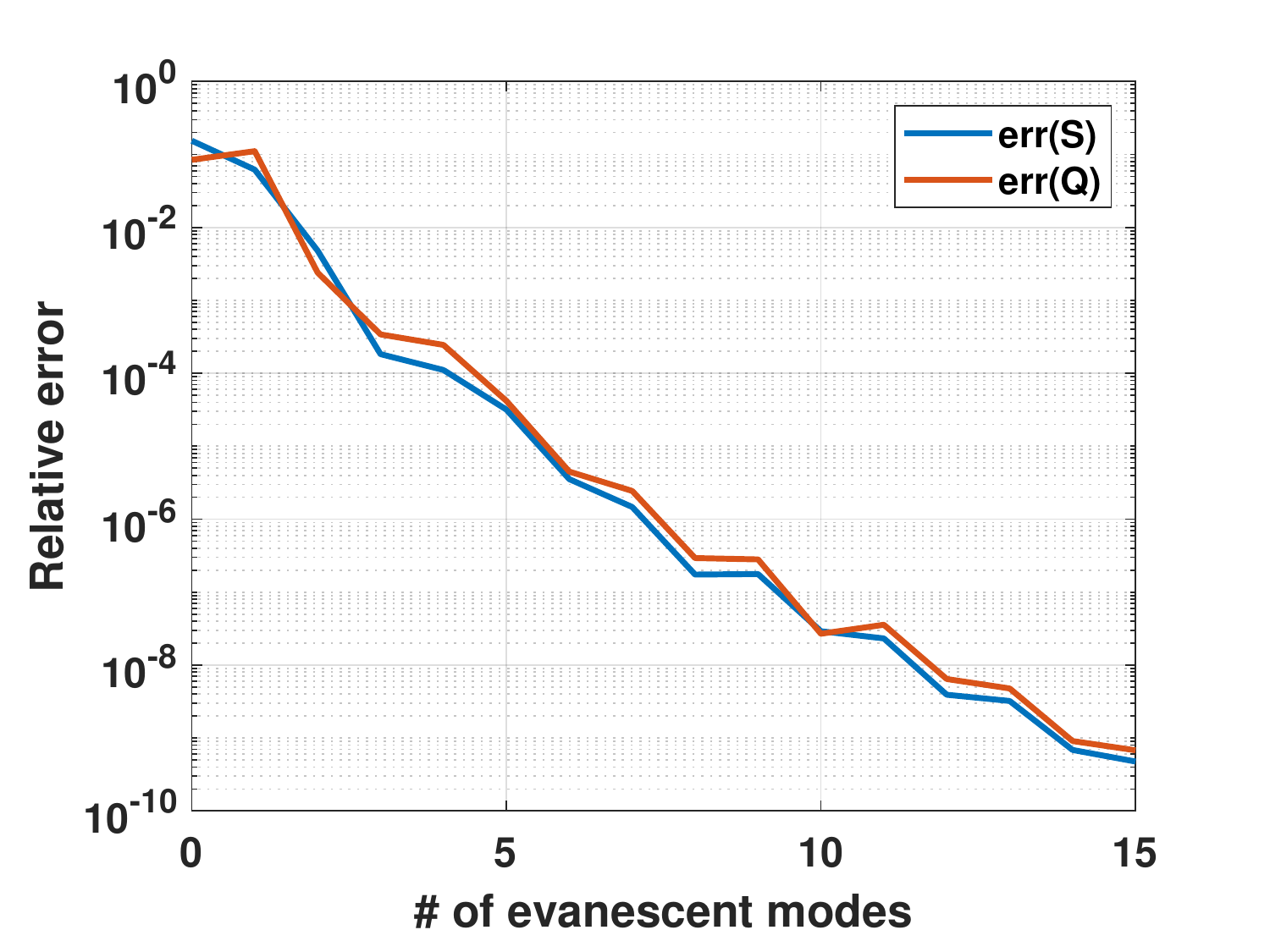}
\caption{Convergence of the cascading procedure as the number of evanescent modes in port \#2 in the guiding system of Fig.~\ref{fig:case2_AB} increases; $err(\matr{S}) = \| \matr{S}_C - \matr{S}_{A+B} \|_F / \| \matr{S}_C \|_F$ and $err(\matr{Q}) = \| \matr{Q}_C - \matr{Q}_{A+B} \|_F / \| \matr{Q}_C \|_F$.}
\label{fig:case2_convg_cscd}
\end{figure}

The WS modes of a composite system often inherit the spatial structure and time delays of those of its subsystems. To illustrate this, the WS modes for systems \textit{A}, \textit{B}, and \textit{C} are computed at $f=5.878 \times 10^{11}$ Hz via diagonalization of these systems' respective $\matr{Q}_\text{prop}$ matrices. Salient features of a subset of the WS modes of each system along with their time delays are discussed next.

\begin{figure}[htbp!]
\centering \subfloat[Mode \#1, 13.25 mm \label{fig:case2A_WS_1}]{\includegraphics[width=0.4\columnwidth]{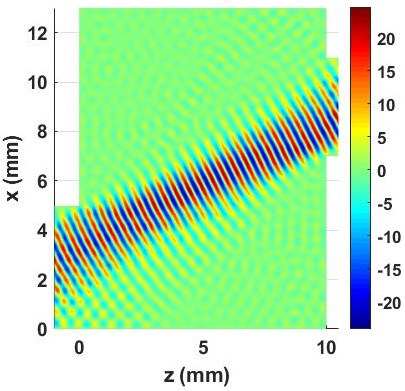}} \qquad
\centering\subfloat[Mode \#10, 22.06 mm \label{fig:case2A_WS_10}]{\includegraphics[width=0.4\columnwidth]{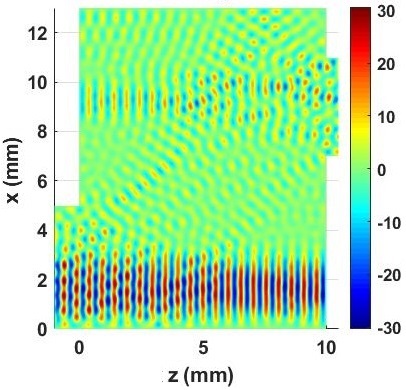}}
\caption{Representative WS modes of subsystem \textit{A}. Distance values below each subfigures are the corresponding equivalent spatial shifts.}
\label{fig:case2A_WSmodes}
\end{figure}

\begin{figure}[htbp!]
\subfloat[Mode \#1, 9.39 mm \label{fig:case2B_WS_1}]{\includegraphics[width=0.5\columnwidth]{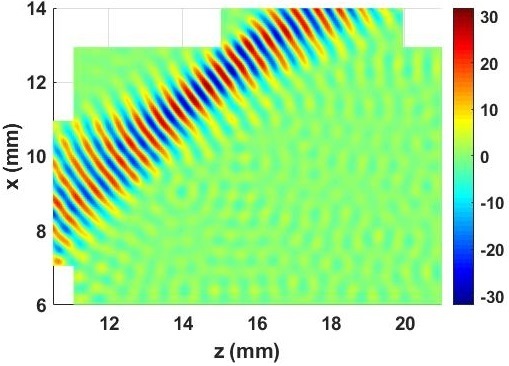}}
\subfloat[Mode \#7, 16.04 mm \label{fig:case2B_WS_7}]{\includegraphics[width=0.5\columnwidth]{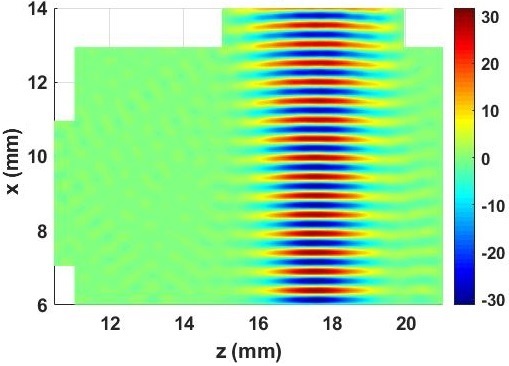}} 
\caption{Representative WS modes of subsystem \textit{B}. Distance values below each subfigures are the corresponding equivalent spatial shifts.}
\label{fig:case2B_WSmodes}
\end{figure}

\begin{figure}[hbtp!]
\subfloat[Mode \#1, 16.04 mm \label{fig:case2C_WS_1}]{\includegraphics[width=0.5\columnwidth]{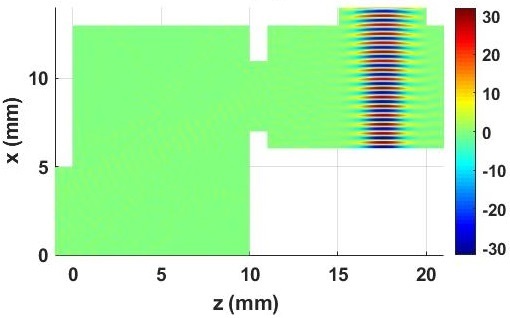}}
\subfloat[Mode \#6, 22.05 mm \label{fig:case2C_WS_6}]{\includegraphics[width=0.5\columnwidth]{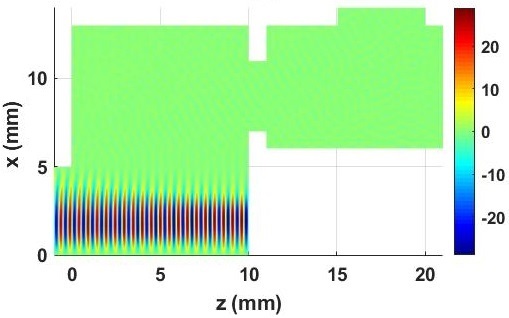}} \\
\centering \subfloat[Mode \#12, 25.11 mm \label{fig:case2C_WS_12}]{\includegraphics[width=0.66\columnwidth]{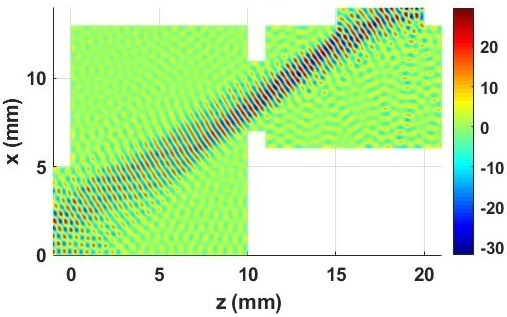}} 
\caption{Representative WS modes of the cascaded system \textit{C}. Distance values below each subfigures are the corresponding equivalent spatial shifts.}
\label{fig:case2C_WSmodes}
\end{figure}

\begin{enumerate}
\item \textit{Subsystem A}. WS mode \#1 enters via port \#1 and exits via port \#2 (or vice versa) without noticeable interaction with the PEC walls; it experiences a time delay that corresponds to the distance between the centers of ports \#1 and \#2 (Fig.~\ref{fig:case2A_WS_1}). WS mode \#10 enters and exits via port \#1 after reflecting off a PEC wall at a nearly normal direction, experiencing a time delay that corresponds to the round-trip between port \#1 and the PEC wall (Fig.~\ref{fig:case2A_WS_10}). 
\item \textit{Subsystem B}. Similar to system \textit{A}, WS mode \#1 is a wave that enters via port \#2 and exits via port \#3 (or vice versa), while WS mode \#7 only excites waves that enter and exit via port \#3 after normally bouncing off a PEC wall (Fig.~\ref{fig:case2B_WSmodes}). 
\item \textit{System C}. Some of the WS modes in system \textit{C} highly resemble those in its subsystems. For example, the spatial structure and time delay of WS mode \#1 of system C (Fig.~\ref{fig:case2C_WS_1}) are nearly identical to those of WS mode \#7 of subsystem \textit{B} (Fig.~\ref{fig:case2B_WS_7}).  Likewise, WS mode \#6 of system \textit{C} (Fig.~\ref{fig:case2C_WS_6}) is very similar to mode \#10 of subsystem \textit{A} (Fig.~\ref{fig:case2A_WS_10}). Other WS modes of system \textit{C} combine traits from those in its subsystems. WS mode \#12, for example, (Fig.~\ref{fig:case2C_WS_12}) enters via port \#1 and exits via port \#3 (or vice versa), and is a hybrid of WS modes \#1 from systems \textit{A} and \textit{B} shown in Figs.~\ref{fig:case2A_WS_1} and \ref{fig:case2B_WS_1}; its spatial shift, to first order, is the sum of those experienced by modes \#1 in systems \textit{A} and \textit{B}. 
\end{enumerate}

\subsection{Periodic Photonic Crystal}

\begin{figure}[hbtp!]
\null \hfill
\subfloat[ \label{fig:case3_PGB1}]{\includegraphics[width=0.7\columnwidth]{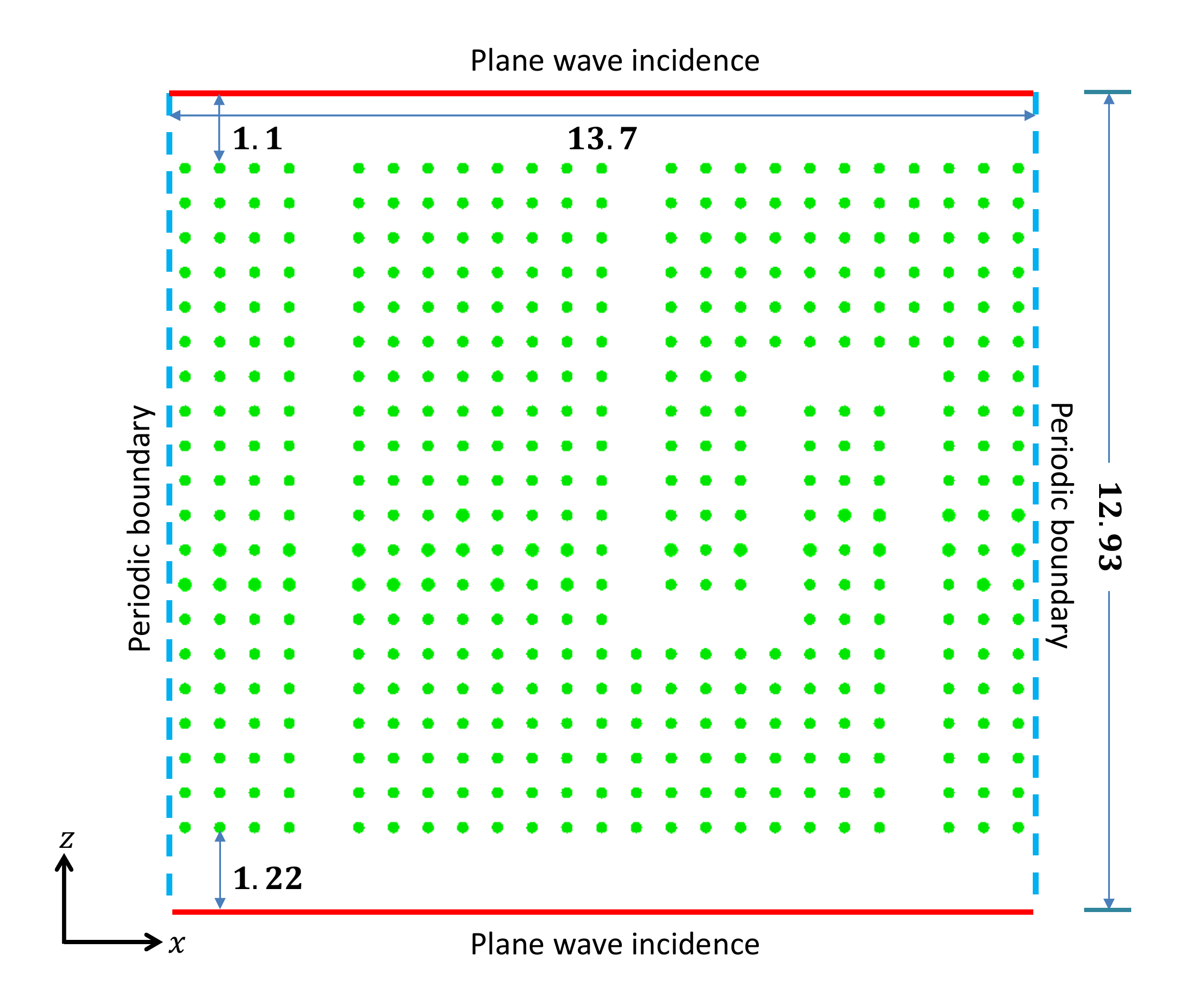}} \hfill
\subfloat[ \label{fig:case3_PGB2}]{\includegraphics[width=0.3\columnwidth]{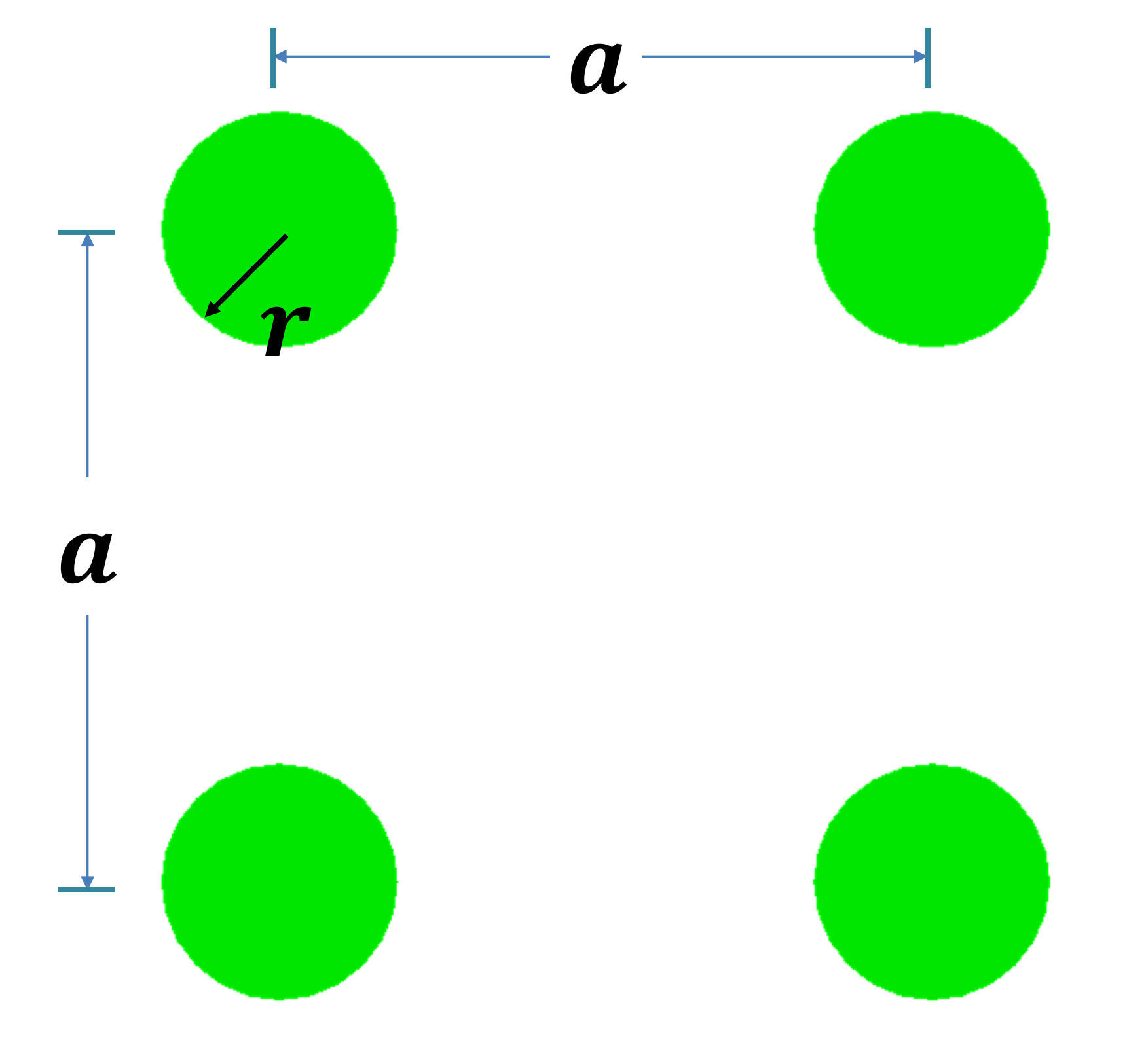}} \hfill
\hfill
\caption{Photonic crystal structure (unit: $\mu$m). The green circles represent silicon rods with relative permittivity $11.56$ embedded in an air background. All rods extend to infinity along the $y$-direction. (a) A unit cell of the system. (b) Local details of rods, $a = 0.548~\mu$m, $r = 0.18a$. Parameters adopted from \cite{Mohammadi2019numerical}.}
\label{fig:case3_PGB}
\end{figure}

Consider the periodic photonic crystal slab composed of unit cells that contain straight and curved waveguides shown in Fig.~\ref{fig:case3_PGB}.  The system is excited by TM$_\text{y}$ fields at $f=1.934 \times 10^{14}$ Hz, which is in the crystal's stop-band.

\begin{figure}[htbp!]
\centering\includegraphics[width=5.0cm]{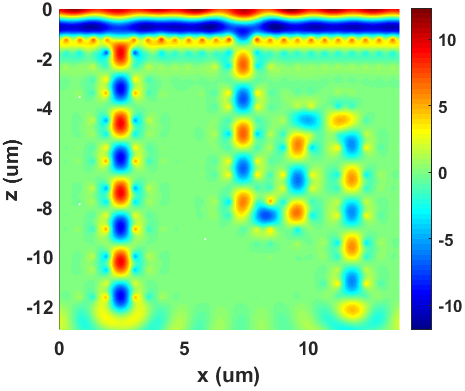}
\caption{Field distribution $real(E_y)$ under a normal plane wave excitation.}
\label{fig:case3_normal}
\end{figure}

First, the slab is excited by a normally incident plane wave. The resulting field distribution is shown in Fig.~\ref{fig:case3_normal} and includes reflected fields as well as relatively weak waves traveling down the waveguides. The reflected and guided waves obviously couple to multiple Floquet modes.

\begin{figure}[htbp!]
\centering\includegraphics[width=6.0cm]{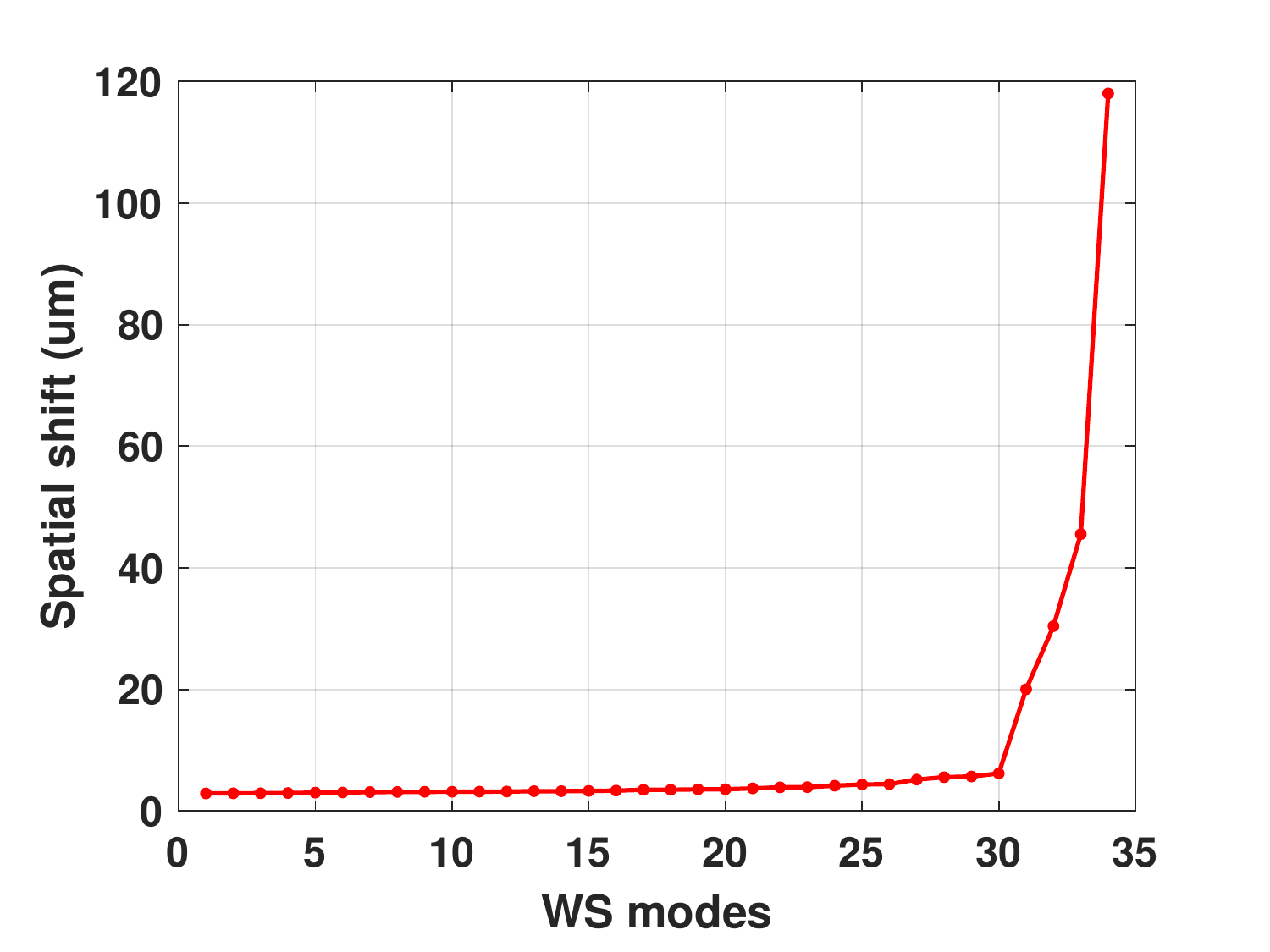}
\caption{Spatial shifts converted from time delays by multiplication with free space light speed, for the photonic crystal structure.}
\label{fig:case3_time_delay}
\end{figure}

\begin{figure}[htbp!]
\null \hfill
\subfloat[Mode \#1, 2.886 $\mu$m \label{fig:case3_WSmode_1}]{\includegraphics[width=0.45\columnwidth]{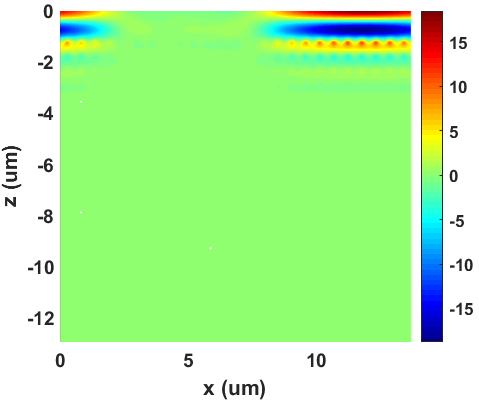}} \hfill
\subfloat[Mode \#28, 5.555 $\mu$m \label{fig:case3_WSmode_28}]{\includegraphics[width=0.45\columnwidth]{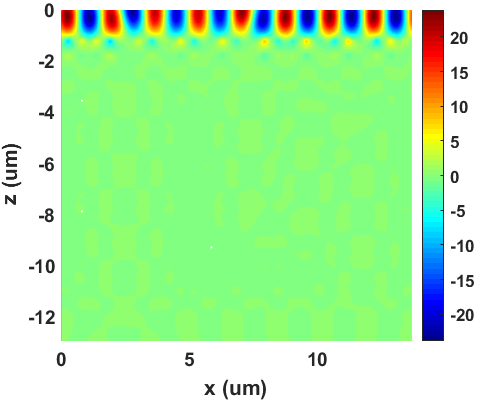}} \hfill
\hfill \null \\
\null \hfill
\subfloat[Mode \#33, 45.55 $\mu$m \label{fig:case3_WSmode_33}]{\includegraphics[width=0.45\columnwidth]{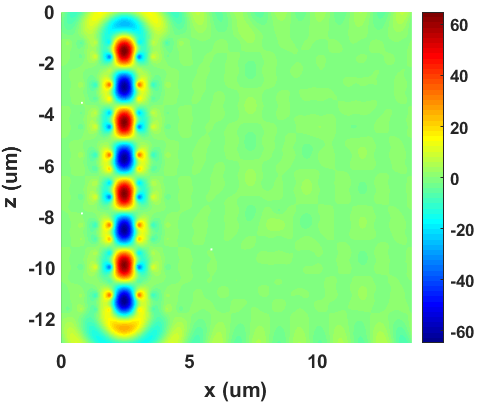}} \hfill
\subfloat[Mode \#34, 118.04 $\mu$m \label{fig:case3_WSmode_34}]{\includegraphics[width=0.45\columnwidth]{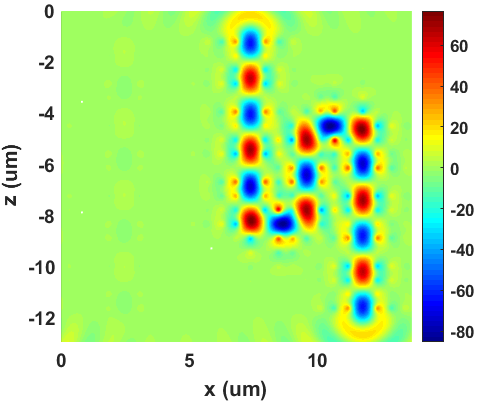}} \hfill
\caption{Several WS modes $real(E_y)$ of the photonic crystal. Distance values below each subfigures are the corresponding equivalent spatial shifts in Fig.~\ref{fig:case3_time_delay}.}
\label{fig:case3_WSmodes}
\end{figure}

Next, the structure's WS time delay matrix $\matr{Q}$ is constructed using a total of $M=74$ modes, or 40 in excess of the number of propagating modes. The WS relationship \eqref{eq:WS_gen2} exhibits an error of $2.7 \times 10^{-6}$. Time delays/spatial shifts and WS modes are constructed by diagonalizing $\matr{Q}_\text{prop}$ and shown in Figs.~\ref{fig:case3_time_delay} and \ref{fig:case3_WSmodes}, respectively. The WS modes belong to one of two categories.

\begin{enumerate}
\item \textit{Reflection modes}. WS modes \#1--\#30 reflect off the top or bottom surface of the crystal, without entering the channels. Mode \#1 (Fig.~\ref{fig:case3_WSmode_1}) has the smallest incident angle, therefore its travel time is the shortest. By comparison, mode \#28 (Fig.~\ref{fig:case3_WSmode_28}) has a larger incident angle, resulting in longer travel time. 
\item \textit{Transmission modes}. WS modes \#33 and \#34 (Figs.~\ref{fig:case3_WSmode_33} \ref{fig:case3_WSmode_34}) are excited by beams targeting the apertures of the straight and curved waveguides, respectively, launching waves that travel down both channels. The spatial shifts experienced by these modes are proportional to the length of the channel in which they travel, and much larger than those of the reflection modes. Modes \#31 and \#32 (not shown) are similar to Modes \#33 and \#34 except that the phases of the fields exciting the apertures on both sides result in reduced energy stored in the slab, and hence smaller time delays. 
\end{enumerate}

\begin{figure}[hbtp!]
\centering\subfloat[ \label{fig:case3_WS_R}]{\includegraphics[width=0.8\columnwidth]{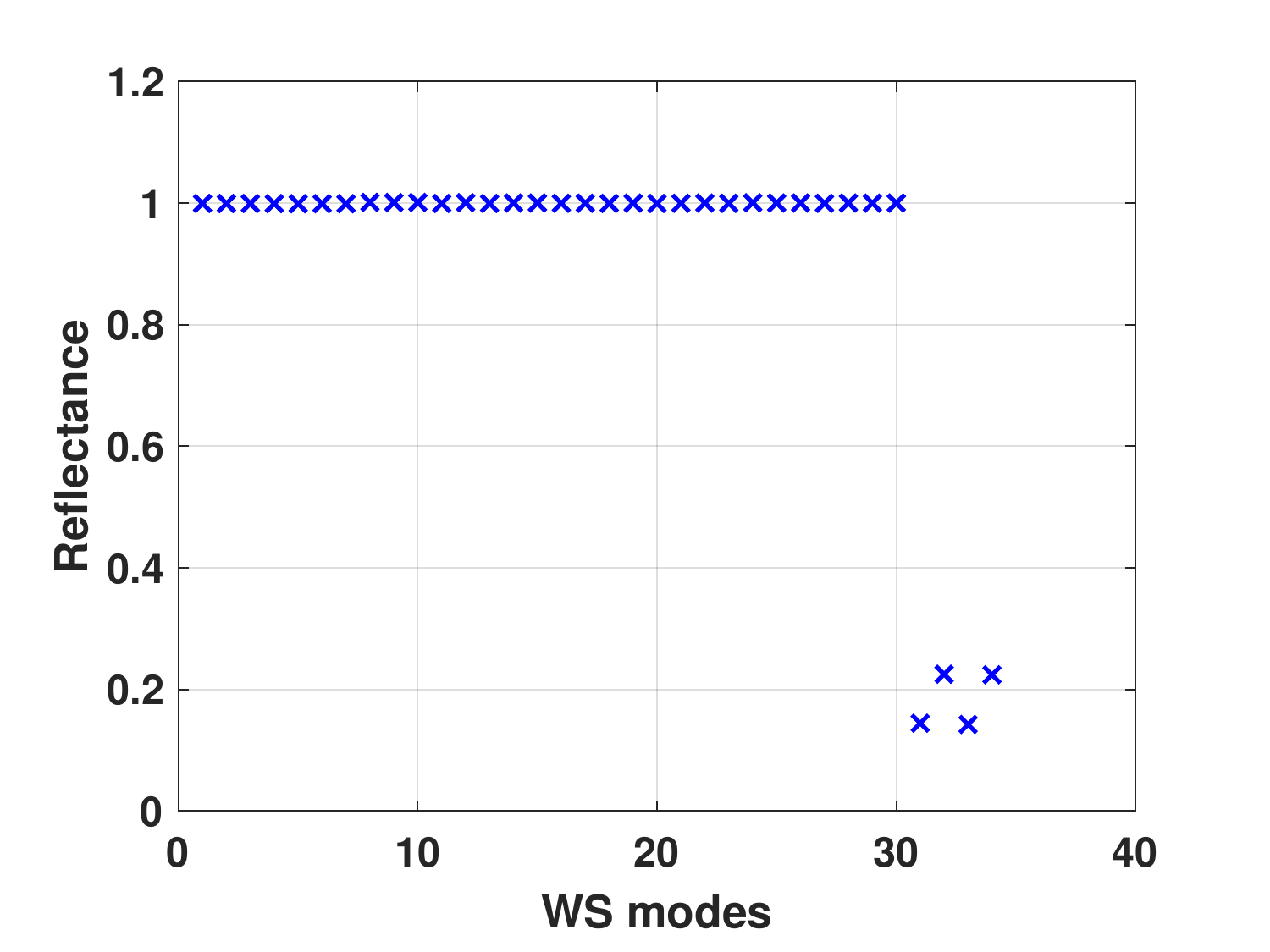}} \\
\centering\subfloat[ \label{fig:case3_Floquet_RT}]{\includegraphics[width=0.8\columnwidth]{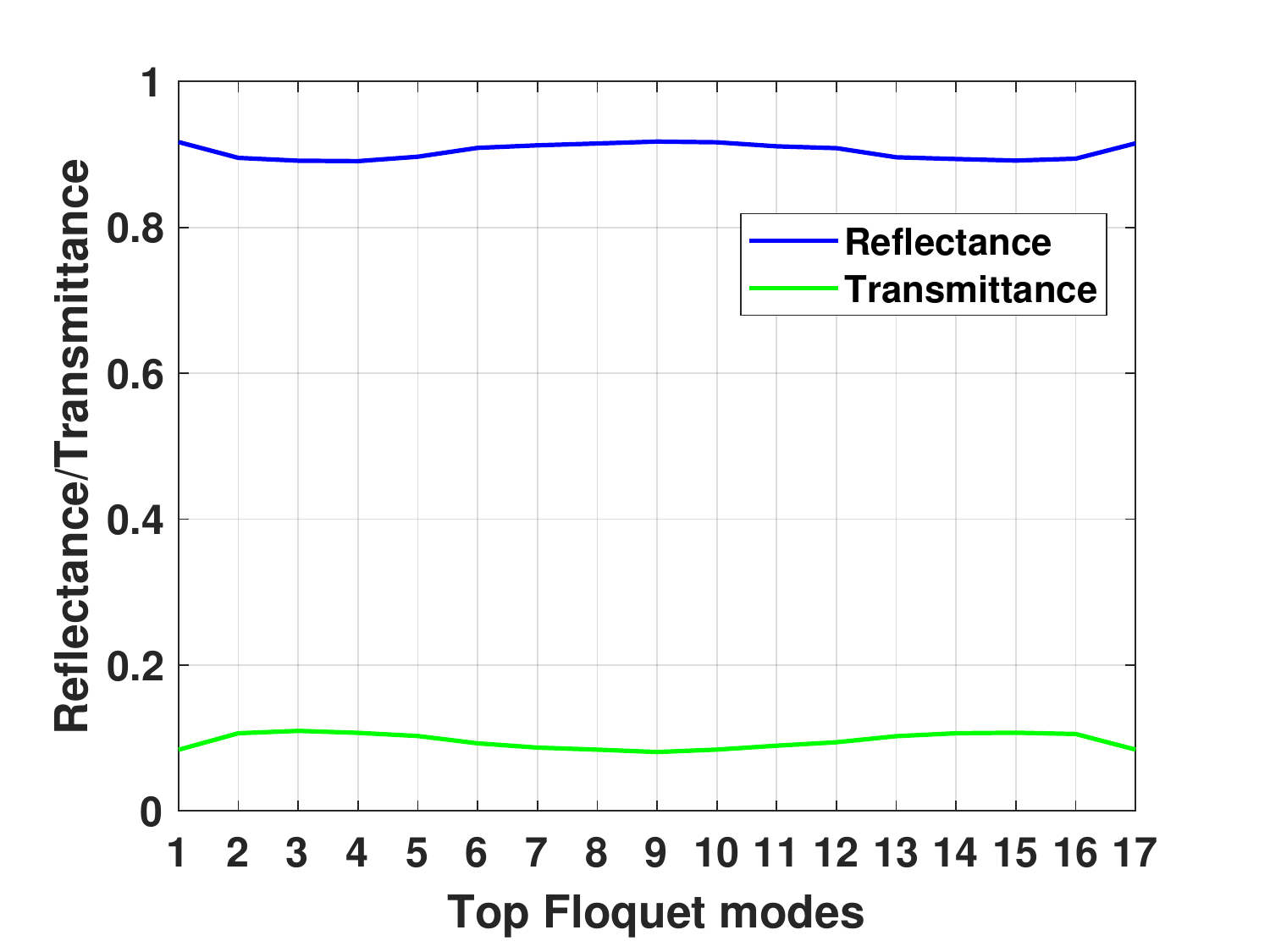}}
\caption{Comparison of global reflectance between Floquet modes and WS modes. (a) Global reflectance obtained by exciting ``incomplete WS modes'' at one of the Floquet ports; (b) Global reflectance/transmittance of propagating Floquet modes at the top port. Floquet mode \#9 corresponds to the normal incidence shown in Fig.~\ref{fig:case3_normal}.}
\label{fig:case3_RT}
\end{figure}

The difference between the reflection and the transmission modes is further illustrated in Fig.~\ref{fig:case3_WS_R}, which shows the fraction of the energy reflected from the periodic system when excited by ``incomplete WS modes'' obtained by retaining only the portion of a WS mode such that illuminates one side. 
For WS modes \#1--\#30, only one of the Floquet ports is strongly excited and chosen as the incident port for reflectance calculation; for WS modes \#31--\#34, both Floquet ports are strongly excited, therefore the reflectance can be calculated by exciting either Floquet port. 
It is evident from Fig.~\ref{fig:case3_WS_R} that WS modes \#1--\#30 excite waves that mostly reflect off the structure, while most energy carried by WS modes \#31--\#34 makes its way through the channel and reaches the other port, with reflectances 14.57\%, 22.45\%, 14.26\% and 22.36\%. 
By contrast, Fig.~\ref{fig:case3_Floquet_RT} shows that all the Floquet modes have reflectance of roughly 90\%, indicating a more average behavior than the WS modes. No Floquet modes can transfer most of its energy through the structure. 

Properties of WS modes can be better understood based on the above observations. Each WS mode only couples to itself and is frequency stable with minimal dispersion. Using WS modes to untangle the physical scattering phenomena, only the \textit{transmission modes} may be excited to transfer the majority of energy through a complicated structure while preserving the information of the signal.

\section{Conclusion}
\label{sec:conclusion}

A generalized WS relationship for guiding and periodic systems with ports that support evanescent modes was developed. The proposed formulation generalizes many previously established WS relationships for systems fed through ports that exclusively support propagating waves. The new formulation not only allows for the application of WS techniques to systems with inhomogeneities near ports, but also allows for the construction of WS time delay matrices of composite systems from those of its potentially tightly coupled subsystems.

Numerical examples involving both guiding and periodic systems validated the proposed formulation and demonstrated the use of WS time delay concepts for constructing WS modes that exhibit well-defined time delays upon interacting with a system through ports supporting evanescent fields.

\appendices

\section{Floquet Modes}
\label{Appdix:Floquet_mode}

Similar to waveguides, fields interacting with periodic structures with normal incidence can be expanded by Floquet modes. The profile for TE$_{\xi}$ modes is
\begin{align}
\bm{\mathcal{X}}_{\text{TE},p}
= \frac{-\hat{\eta} k_{\zeta n} + \hat{\zeta} k_{\eta m}}{\sqrt{L_\eta L_\zeta (k_{\eta m}^2 + k_{\zeta n}^2)}} e^{-j k_{\eta m} \eta} e^{-j k_{\zeta n} \zeta}
\end{align}
and that for TM$_{\xi}$ modes is
\begin{align}
\bm{\mathcal{X}}_{\text{TM},p}
= \frac{\hat{\eta} k_{\eta m} + \hat{\zeta} k_{\zeta n}}{\sqrt{L_\eta L_\zeta (k_{\eta m}^2 + k_{\zeta n}^2)}} e^{-j k_{\eta m} \eta} e^{-j k_{\zeta n} \zeta} \,.
\end{align}
Here $p$ maps to the tuple $(m, n)$; $L_\eta$ and $L_\zeta$ are periodicities of the unit cells; and $k_{\eta m} = - \frac{2 \pi}{L_\eta} m$, $k_{\zeta n} = - \frac{2 \pi}{L_\zeta} n$.
The propagation constant in $\xi$ of the $p$-th Floquet mode is
\begin{align}
\beta_p = \sqrt{k^2 - k_{\eta m}^2 - k_{\zeta n}^2} \,.
\end{align}
The mode impedance is $Z_p = k Z / \beta_{p}$ for TE$_{\xi}$ modes, and $Z_p = \beta_{p} Z / k$ for TM$_{\xi}$ modes, where $Z = \sqrt{\mu / \varepsilon}$ at the Floquet port.

\ifCLASSOPTIONcaptionsoff
  \newpage
\fi

\bibliographystyle{IEEEtran}
\bibliography{IEEEabrv,ref}

\renewenvironment{IEEEbiography}[1]
  {\IEEEbiographynophoto{#1}}
  {\endIEEEbiographynophoto}

\end{document}